\documentclass[twocolumn,nofootinbib,superscriptaddress,preprintnumbers,floatfix]{revtex4}

\usepackage[hyperfootnotes=false]{hyperref}
\usepackage{graphicx}
\usepackage[small]{subfigure}
\usepackage{amsmath}
\usepackage{xspace}
\usepackage{color}

\newcommand{\eq}[1]{Eq.~\eqref{eq:#1}}

\renewcommand{\sec}[1]{Sec.~\ref{sec:#1}}

\newcommand{\subsec}[1]{Sec.~\ref{subsec:#1}}
\newcommand{\fig}[1]{Fig.~\ref{fig:#1}}

\newcommand{\be}[0]{\begin{equation}}
\newcommand{\ee}[0]{\end{equation}}

\newcommand{\ord}[1]{\mathcal{O}(#1)}

\newcommand{\sfrac}[2]{\textstyle{\frac{#1}{#2}}}

\newcommand{\bn}{\bar{n}}
\newcommand{\kt}{\textrm{k}_\textrm{T}}

\newcommand{\ket}[1]{\vert #1 \rangle}
\newcommand{\R}{\widehat{\ca{R}}}

\newcommand{\ca}[1]{\mathcal{#1}}

\newcommand{\SCETI}{SCET$_{\rm I}$\xspace}
\newcommand{\SCETII}{SCET$_{\rm II}$\xspace}
\newcommand{\SCETp}{SCET$_{+}$\xspace}
\newcommand{\Lqcd}{\Lambda_{\rm QCD}\xspace}

\newcommand{\nn}{\nonumber}

\begin{document}


\preprint{\vbox{\hbox{INT-PUB-11-050} } }

\title{Factorization Constraints on Jet Substructure}

\author{Jonathan R. Walsh and Saba Zuberi}
\affiliation{Theoretical Physics Group, Ernest Orlando Lawrence Berkeley National Laboratory, and Center for Theoretical Physics, University of California, Berkeley, CA, 94720, USA\vspace{2ex}}

\begin{abstract}
Factorization underpins our ability to make predictions at the LHC, both in Monte Carlo simulations and direct calculations.  An improved theoretical understanding of jet substructure can lead to calculations that can confront data and validate the Monte Carlo description of jets.  We derive constraints on jet substructure algorithms from factorization, focusing on the broad class of jet observables where the soft and collinear dynamics of QCD dominate.  A necessary condition for factorization is that the phase space constraints on soft and collinear dynamics for a given observable are independent of each other. This condition allows us to use a simple power counting analysis to place strong constraints on the form of jet substructure methods that can factorize. We illustrate this approach by considering four substructure algorithms, the mass-drop filter Higgs tagger, pruning, trimming, and $N$-subjettiness. We find that factorization constrains the scaling of parameters in all of the substructure algorithms, and that the generic declustering and filtering techniques do not factorize.  However, we determine simple modifications that can be made to allow for factorization.
\end{abstract}

\maketitle

\tableofcontents

\section{Introduction}
\label{sec:intro}

Jets play a central role in the search for new physics at hadron colliders. Jet algorithms allow us to map complicated hadronic final states onto the underlying hard interaction and therefore are important in both finding new physics signals and understanding the dominant backgrounds from QCD.

At the new energy frontier of the LHC, heavy particles such as electro-weak bosons are often produced with sufficient boost that their hadronic decay products are reconstructed in to a single ``fat jet" using traditional jet algorithms. This makes their identification challenging. There has been much recent theoretical progress in developing jet substructure techniques, which have two goals. First, they use the high $p_T$ substructure in the jets to improve the discrimination between jets from the decay of heavy boosted objects and large $p_T$ ``QCD jets" from the QCD evolution of a colored parton. Second, they also aim to remove the contamination of soft radiation from underlying event and pileup, which is important to improve the resolution of measurements for large jets.

Jet substructure tools can extend the reach of a variety of SM analyses and provide new search strategies for BSM physics.  Substructure algorithms have been developed for specific channels, such as top-tagging~\cite{Butterworth:2002tt, Thaler:2008ju, Kaplan:2008ie, Almeida:2008tp, Plehn:2010st, Rehermann:2010vq, Jankowiak:2011qa, Thaler:2010tr, Thaler:2011gf}, $W$-tagging \cite{Seymour:1993mx,Butterworth:2002tt,Cui:2010km}, Higgs + $W/Z$ production, resurrecting the $h \to b \bar{b}$ search channel that would otherwise be swamped by the $W/Z$ + jets background~\cite{Butterworth:2008iy, Plehn:2009rk, Soper:2010xk}, and new physics channels \cite{Butterworth:2007ke, Butterworth:2009qa, Kribs:2009yh, Bhattacherjee:2010za, Kribs:2010hp, Kribs:2010ii, Hackstein:2010wk, Englert:2010ud, Chen:2010wk, Katz:2010iq, Katz:2010mr, Plehn:2011tf}. Substructure algorithms have also been developed for generic boosted topologies, which are referred to as jet grooming~\cite{Ellis:2009su, Ellis:2009me, Krohn:2009th}.  These methods have been shown to be broadly effective at tagging boosted decays and removing contamination from underlying event and pileup.  Jet shapes, which measure a function of the momenta within a jet without placing strong exclusive cuts on the phase space have recently been shown to be effective in tagging boosted tops \cite{Almeida:2008tp, Jankowiak:2011qa,Thaler:2010tr,Thaler:2011gf}.  By avoiding the rather complex and exclusive phase space cuts ``traditionally'' imposed by jet substructure techniques, jet shapes are generally simpler and more amenable to direct calculation.

Many jet substructure algorithms are beginning to be verified on data~\cite{Miller:2011qg, Rappoccio:2011nj, Aaltonen:2011pg} and as with other jet tools, it is important to understand them from a theoretical perspective.  Currently we rely on Monte Carlo (MC) predictions for jet substructure, which are relatively untested and may be susceptible to large, poorly quantified uncertainties.  The MC lacks perturbative accuracy in describing the evolution of the jet, which is compensated by good modeling of hadronization and non-perturbative effects.  Calculations of jet substructure can be compared to data and help validate and improve the MC description of jet substructure. Furthermore, the plethora of available jet substructure techniques poses a challenge in exploring them experimentally. The Boost 2010 Monte Carlo study~\cite{Abdesselam:2010pt} suggests that many methods have similar quantitative performance, meaning it is important to identify any characteristics of substructure algorithms that make them theoretically attractive.

Our ability to make reliable theoretical predictions at the LHC relies on factorization, which is the separation of the cross section into pieces that depend on separate energy scales. There exists a hierarchy of perturbative scales in most jet physics applications, i.e., the jet $p_T$ of order $\sqrt{\hat{s}}$, the invariant mass of the jet, $m_J$, which is typically much smaller.  In this case the soft and collinear limits of QCD provide the correct description of jet evolution, where the collinear dynamics generates high energy radiation collimated with the jet direction and the soft dynamics governs much lower energy global radiation in and between jets.  This gives rise to large logarithms in a fixed order perturbative description of a jet that must be resummed to control the perturbative series and regain accuracy. Factorization is necessary not just in separating perturbative and non-perturbative effects, but also in disentangling the scales set by the dynamics of soft and collinear evolution that describe a jet, see \fig{Scales}. This is essential for resummation of large logs carried out by renormalization group evolution.

This work examines a necessary condition for factorization for jet observables. We focus on soft-collinear factorization, the separation of the soft and collinear contributions to a jet observable, which is a key step in proving factorization theorems for a wide range of jet physics applications, including most jet algorithms, substructure methods, and jet shape observables. A necessary condition for soft-collinear factorization is that the phase space cuts generated by the observable being measured constrain the collinear dynamics independently of the soft radiation, and vice versa to all orders in perturbation theory. We refer to this as soft-collinear observable phase space (OPS) factorization.

We find that a straightforward\footnote{By straightforward we mean that it can be performed without technical expertise in SCET.} power counting analysis can be used to test whether OPS factorization holds for a given jet observable by using Soft Collinear Effective Theory (SCET)~\cite{Bauer:2000ew, Bauer:2000yr,Bauer:2001ct, Bauer:2001yt}, an effective field theory (EFT) of QCD at high energies. The EFT framework is very useful because it provides a systematic power counting which allows one to quantitatively determine the soft and collinear contributions to a jet observable and the corrections to this limit.  SCET has been applied to a broad range of jet physics problems, many relevant for jet substructure.  It has been used to understand $e^+e^-$ event shapes at high accuracy~\cite{Schwartz:2007ib,Bauer:2008dt,Becher:2008cf,Hornig:2009vb,Abbate:2010xh}, jet algorithms and jet shapes in multijet events~\cite{Cheung:2009sg,Ellis:2009wj,Ellis:2010rwa,Kelley:2011tj}, and ways to use event shapes to veto against jet-like events at the LHC~\cite{Stewart:2010tn,Berger:2010xi,Jouttenus:2011wh}.  The framework of SCET can be used to analyze whether the phase space cuts generated by the observable that govern the soft and collinear dynamics factorize.  It allows one to easily determine the key qualitative features of the algorithm or observable, which we find is informative, especially for jet substructure.  When factorization fails, this analysis can be used to diagnose problems and identify modifications that allow for factorization. The analysis of OPS factorization cannot by itself prove factorization, but is a key component in the proof.

Jet algorithms, substructure methods, or jet shapes for which soft-collinear phase space constraints do not separate cannot be factorized.  If large logarithms arise in a fixed order perturbative series, then although the perturbative series is calculable, these large logs cannot be resummed and accurate predictions cannot be made.  This has impact not only on a calculation but on the MC as well, as factorization underlies the MC.  If factorization is not respected for the leading logarithms, then it endangers the predictability of the MC.

The outline of the paper is as follows: in \sec{fact}, we discuss modes and power counting in SCET, followed by a review of factorization in the effective theory.  In \sec{tests}, we show how we can use power counting of soft and collinear modes in SCET to test for OPS factorization, and we present general constraints that apply to jet substructure methods.  We illustrate the power counting analysis of OPS factorization by applying it to the $\kt$ class of jet algorithms ($\kt$, Cambridge/Aachen, and anti-$\kt$) \cite{Catani:1991hj,Catani:1993hr,Ellis:1993tq,Dokshitzer:1997in,Cacciari:2008gp}, as well as the JADE algorithm \cite{Bartel:1986ua,Bethke:1988zc}, which is known not to factorize.  The power counting analysis also lets us easily determine characteristic properties of each jet algorithm.  In \sec{jetsub}, we apply our test to four substructure algorithms: the mass-drop filter Higgs tagger, pruning, and trimming, in addition to the jet shape $N$-subjettiness.  We find that the general procedures of declustering and filtering used in the Higgs tagger do not factorize.  We discuss simple modifications that can restore factorization, and study the impact on the effectiveness of the Higgs tagger. We find that both pruning and trimming factorize if the parameters of the algorithms are chosen to have the appropriate scaling with the power counting parameter of SCET.  We use the power counting analysis to build a simple picture for the behavior of pruning, and we find this picture agrees well with Monte Carlo simulation.  We discuss factorization constraints for generic jet shapes and $N$-subjettiness in particular.  Finally, in \sec{conclusions} we conclude.

\section{Factorization and Power Counting}
\label{sec:fact}

In this section we highlight the distinction between the factorization of a hard scattering process and soft-collinear factorization, which is required to obtain a reliable perturbative description when there exists a hierarchy of scales in a jet observable. Factorization requires that the phase space constraints generated by the observable act on soft and collinear final states independently of each other. We refer to this as OPS factorization. Power counting in SCET makes it straightforward to test this necessary condition for soft-collinear factorization. We review SCET power counting and illustrate the steps needed to prove factorization with the example of the dijet jet mass distribution. 

\subsection{Soft Collinear Factorization}
\label{subsec:SCfact}

Hard scattering processes at hadron colliders with no constraints on the final state satisfy a familiar factorization:
\be
\label{eq:hardfact}
\sigma = \sum_{i,j} \int dx_i \, dx_j \, f_i (x_i) f_j (x_j) \, \widehat{\sigma}_{ij} (x_i,x_j) \,,
\ee
where $\widehat{\sigma}_{ij}$ is the hard scattering cross section that has been factorized from the parton distribution functions $f_{i,j}$.  This factorization separates the calculable, perturbative hard scattering at the center-of-mass energy scale, $\sqrt{\hat{s}}$, from the measurable non-perturbative parton luminosities at the scale $\Lqcd$.  \eq{hardfact} reflects the basic action of factorization: it divides the cross section into separately calculable or measurable pieces that each depend on dynamics at different scales.  This basic factorization theorem is widely applied to make theoretical predictions for hadron collisions.  For example, it is part of the underlying assumptions used in MC programs to generate events.

The utility of the hard scattering factorization in \eq{hardfact} is limited, since it applies only to fully inclusive cross sections.  In all jet physics applications the cross section is made more exclusive through final state cuts and the factorization theorem becomes more complex, if it holds at all.  It is important to identify which applications fail factorization, since it greatly compromises our ability to make reliable theoretical predictions for such observables and therefore limits their utility.

For the jet mass, there is a natural hierarchy of scales in a jet between the jet $p_T$, the invariant mass of the jet, $m_J$, and the non-perturbative scale $\Lqcd$, where $p_T \gg m_J \gg \Lqcd$.  The perturbative scales associated with the soft and collinear dynamics governing the jet mass are shown in \fig{Scales}. This hierarchy of scales can lead to large Sudakov double logarithms, giving terms like $\alpha_s^n \ln^{2n} m_J/p_T$ in perturbation theory. This can spoil the convergence of the perturbative expansion unless these logarithms are resummed to all orders.

Restoring perturbative control requires the hard factorization in \eq{hardfact} to be extended by showing soft-collinear factorization. When soft-collinear factorization can be shown, the non-perturbative contribution of the hadronic initial state is factorized from the perturbative components, as in \eq{hardfact}, while $\hat{\sigma}_{ij}$ is further factorized by separating the collinear dynamics at the scale $m_J$ from global soft radiation at the scale $m_J^2/p_T$.

SCET provides a systematic framework to carry out soft-collinear factorization in the appropriate limit of QCD. For a $pp \to N$-jet observable, such a factorization theorem takes the schematic form~\cite{Bauer:2002nz, Bauer:2008jx, Stewart:2009yx, Ellis:2010rwa, Stewart:2010tn}
\be \label{eq:SCETfact}
\sigma_N = H_N \left[ B_a \times B_b \times \prod_{k=1}^{N} J_k \right] \, \otimes \, S_N \,.
\ee
The hard function $H_N$ comes from the short-distance hard scattering process, while the beam functions $B$ and jet functions $J$ come from the collinear evolution of the initial and final hard partons from the hard scattering.  The beam functions have an internal factorization containing the parton distribution functions.  Finally, the soft function $S_N$ comes from global soft radiation.  The presence of the additional components in the factorization theorem in \eq{SCETfact} compared to \eq{hardfact} reflects the fact that new scales are introduced in to the problem by the final state cuts which define the jets.  When factorization can be shown, OPS factorization is the physical reason that we can separate the collinear and soft evolution of the jets.  It states that the soft and collinear contributions to the measurement are independent.  Such a factorization theorem is neither guaranteed nor assumed, as is often the case for hard factorization in \eq{hardfact}.

\subsection{Power Counting in SCET}
\label{subsec:SCEToverview}

\begin{figure}[t]
\centerline{ \scalebox{0.32}{\includegraphics{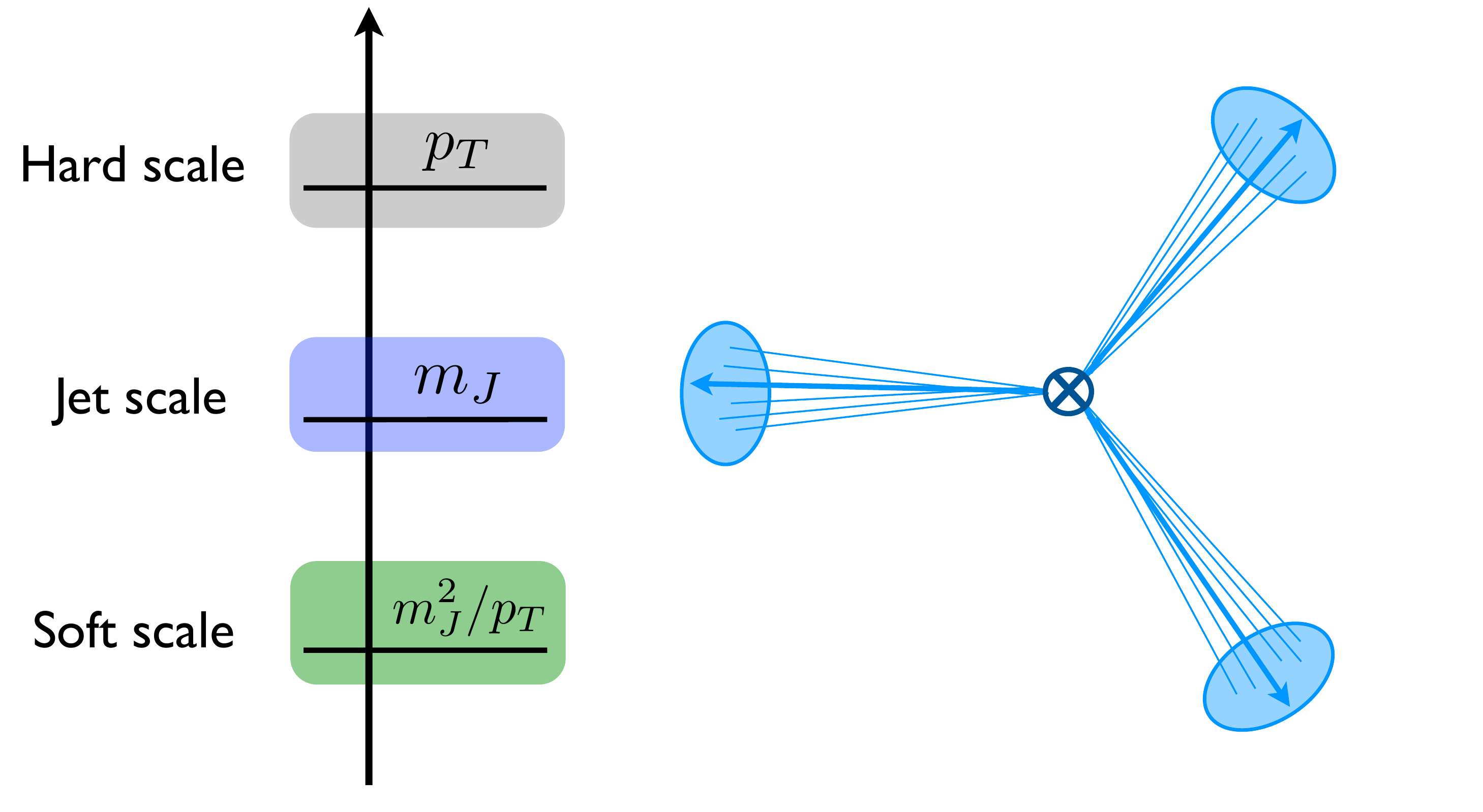}} }
\caption{Hierarchy of scales relevant for widely separated collimated jets.}
\label{fig:Scales}
\end{figure}

Soft and collinear radiation in QCD provides the dominant description of jet evolution in many jet physics applications. In these cases there is typically a hierarchy of scales related to the observable.  The effective field theory SCET provides a systematic expansion around the soft-collinear limit of QCD below a hard scale, which simplifies the proof of soft-collinear factorization. This is done by expanding in a power counting parameter $\lambda \ll 1$ and gives rise to soft and collinear modes, which become fields in SCET. For jet mass, $\lambda = m_J/E_J$.  Their momentum scaling, expressed in terms of light cone components  $p^\mu = (n \cdot p, \bn \cdot p, p_\perp) \equiv (p^+,p^-,p_\perp)$, is\footnote{Any four-vector can be decomposed as
\begin{align}
p^\mu =  n\cdot p \frac{ \bn^\mu}{2}+ \bn \cdot p \frac{n^\mu}{2}+p_\perp^\mu \, , \nn
\end{align} 
and $p^2=p^+ p^- + p_\perp^2$.}  
\begin{align} \label{eq:scScales}
\textrm{collinear: } & p_c = (p_c^+, p_c^-, p_c^{\perp}) \sim E_{J}(\lambda^2, 1, \lambda) \,, \nn \\
\textrm{soft: } & p_s = (p_s^+, p_s^-, p_s^{\perp}) \sim E_{J}(\lambda^2, \lambda^2, \lambda^2) \,,
\end{align}
where $n$ and $\bn$ are light-cone vectors, $n^{\mu} = (1,\hat{n})$ and $\bn^{\mu} = (1,-\hat{n})$, with $\hat{n}^2 =1$.  The jets shown in \fig{Scales} have separate collinear modes, with momentum scaling defined with respect to the jet direction $n_i$.  The exchange of collinear radiation between \textit{different} jet directions has invariant mass of order the hard scale $p_T \sim E_J$ and is integrated out of the effective theory, but collinear radiation inside of a jet remains.  We note that the large component of collinear modes scales as the jet energy, so the relevant logs are $m_J / E_J$.  For light jets, $E_J = p_T \cosh y$, and the pseudorapidity factor can enhance logs of $m_J/p_T$.  Soft modes, which have homogenous scaling, can be exchanged between different collinear sectors.

The momentum scaling in \eq{scScales} simplifies the physical picture of jet events.  This is seen from the kinematics of soft and collinear modes.  The relative scaling of their energies is
\be \label{eq:scEratio}
\frac{E_{s}}{E_{c}} \sim \lambda^2 \,.
\ee
Therefore the energy, or equivalently transverse momentum, of the jet  is determined by the collinear particles in the jet, up to power corrections of $\ord{\lambda^2}$ from soft particles. The angle, $\Delta R^2 = \Delta \eta^2 + \Delta \phi^2$, between collinear particles in the same jet (collinear sector) is
\be \label{eq:ccangle}
\Delta R_{c_1 c_2} \sim \lambda \,.
\ee
The angle between soft particles, which interact globally in the event, is $\ca{O}(1)$:
\be \label{eq:ssangle}
\Delta R_{s_1 s_2} \sim 1 \,.
\ee
Finally we can use power counting to show that the angle between a soft particle and any collinear particle in a jet is the same at leading power and is equal to the angle of the soft particle to the jet axis. For the polar angle $\theta_{cs}$, the relation is
\begin{align} \label{eq:nspolarangle}
\frac{2p_c \cdot p_s}{E_c E_s} &= 2(1-\cos\theta_{cs}) = \frac{4p_s^+}{p_s^+ + p_s^-} + \ca{O}(\lambda^2) \nn \\
&= 2(1-\cos\theta_{ns}) + \ca{O}(\lambda^2) \,.
\end{align}
An analogous relation holds for $\Delta R$:
\be \label{eq:nsangle}
\Delta R_{cs} = \Delta R_{ns} + \ca{O}(\lambda^2) \,.
\ee
The angle $\Delta R_{ns}$ scales as $\lambda^0$.  Soft particles do not distinguish the individual collinear splittings in a jet, but instead see each collinear sector as a color source moving along the light-like direction $n_i$. The combination of a soft and collinear particle in a jet, $p_c + p_s$, only changes the $\ca{O}(\lambda^2)$ component of the collinear momentum:
\be \label{eq:softindep}
p_c + p_s = (p_c^-, p_c^+ + p_s^+, p_c^{\perp})[1+\ca{O}(\lambda)] \,.
\ee
Since the $p_c^-$ and $p_c^{\perp}$ components determine the energy and direction of the collinear mode, these are unchanged up to power corrections.  

The explicit scaling of soft and collinear modes lets us construct a simple physical picture of the jet and gives us a framework to explore jet properties.  Collinear particles are restricted to a narrow region of angle $\sim\lambda$ around the jet axis, and carry the total energy of the jet up to power corrections in $\lambda$.  Soft particles populate the entire jet, and they cannot resolve individual collinear particles.  Instead, soft particles only resolve the jet direction.  The measurement can add dependence on the total energy of each collinear sector.  This basic picture can let us study the properties of different jet algorithms, which we do in \subsec{jetalgs}.

It is important to note that the scaling of the kinematics of soft and collinear modes is a \textit{characteristic} scaling. Soft and collinear particles can contribute to the jet cross section at leading power in regions of phase space when their momenta is parametrically smaller than the characteristic scaling in \eq{scScales} (namely when integrating over phase space in real and virtual corrections). For example, the characteristic scaling of the angle between collinear particles is given in \eq{ccangle}, but can also contribute to the jet observable for much smaller values of $\ca{O}(\lambda^2)$. We take this in to account when necessary in our analysis of jet substructure methods. Notice that if the large component of the collinear momentum, $p^-_c$ in \eq{scScales}, becomes $\ca{O}(\lambda^2)$, the collinear and soft modes overlap. This double-counting is explicitly removed from the effective theory by the zero-bin subtraction.  In the next section, we see how power counting is applied to prove soft-collinear factorization.

\subsection{Factorization in SCET}
\label{subsec:factoverview}

Factorization requires proving that the cross section can be separated to all orders in perturbation theory into matrix elements that depend on dynamics at different scales.  In SCET, factorization will express the cross section in a form like \eq{SCETfact}, with independently calculable hard, jet, and soft functions.  Although formal factorization proofs require a significant amount of technical machinery, a simple physical statement is at the heart of soft-collinear factorization.  In SCET, this statement is that the soft and collinear contributions to the observable are distinct at all orders.  A formal factorization proof requires four essential steps:
\begin{list}{\labelitemi}{\leftmargin=1.5em}
\item[1.] We must show that soft and collinear dynamics are the relevant modes to describe the final states and observable we are interested in.  In particular, the scales in the problem are such that SCET with the modes in \eq{scScales} (formally known as \SCETI)\footnote{\SCETII is a formulation of SCET with soft modes whose momenta scale as $p_s \sim p_c^- (\lambda,\lambda,\lambda)$.  Since these modes are parametrically more energetic than the soft modes in \eq{scScales}, soft modes in \SCETI are often referred to as ultrasoft modes.} provides the correct EFT description.
\item[2.] Interactions between soft and collinear fields in the SCET $N$-jet operator must not exist at leading order in $\lambda$.  This soft-collinear decoupling is process independent and follows straightforwardly in the construction of SCET \cite{Bauer:2001yt}.
\item[3.] For the observable of interest, the phase space constraints in the soft and collinear sectors must be independent of each other to all orders in perturbation theory.  We refer to this as soft-collinear \textbf{observable phase space factorization} (OPS), and the focus of this paper is to show that a simple analysis can determine if it is satisfied.
\item[4.] An operator definition must be supplied for all components in the factorization theorem.  This includes a measurement operator which imposes the final state cuts in the factorized cross section.
\end{list}

We discuss these points with an example, $e^+e^- \to 2$ hemisphere jets at center of mass energy $Q$ where we measure the mass $m_{1,2}$ in each hemisphere.  In the limit of small jet mass, soft and collinear modes describe each jet and \SCETI can be used to sum the logs of $m_{1,2} / Q$~\cite{Fleming:2007qr, Fleming:2007xt, Hoang:2008fs, Kelley:2011tj, Hornig:2011iu}.  Up to power corrections in $m_{1,2}^2 / Q^2$, each jet mass is a simple sum,
\be \label{eq:Mjet}
m_J^2 = Q \left( \sum_{c_i \in J} n\cdot p_i^c + \sum_{s_i \in J} n\cdot p_i^s \right) = m_c^2 +m_s^2 \,,
\ee
where $m_{c(s)}$ is the collinear (soft) contribution to the jet mass. The scales associated with the jet and soft functions can be determined by the collinear and soft contributions to the observable.  In \SCETI, the scales in terms of the power counting parameter are
\be
\mu_J \sim Q\lambda \,, \qquad \mu_S \sim Q\lambda^2 \,.
\ee
The contributions to the observable then give the scales $\mu_{J,S}$ in terms of $m_{1,2}$:
\begin{align} \label{eq:massscales}
m_c^2 &= Q \, n\cdot p_c \sim Q^2 \lambda^2 \quad \Rightarrow \quad \mu_J \sim m_{1,2} \,, \nn \\
m_s^2 &= Q \,  n\cdot p_s \sim Q^2 \lambda^2 \quad \Rightarrow \quad \mu_S \sim m_{1,2}^2 /Q \,.
\end{align}
The QCD cross section can be written as a forward scattering matrix element of the current:
\begin{align}
\sigma(m_1, m_2) \sim \langle 0 \lvert \, j_{\rm QCD}^{\dagger} \, \R(m_1,m_2) \,  j_{\rm QCD} \, \rvert 0 \rangle \,. \nn
\end{align}
The restriction operator $\R (m_1, m_2)$ imposes the measurement on the final state,
\be
\R (m_1, m_2) = \delta(m_1 - \widehat{m}_1) \delta(m_2 - \widehat{m}_2) \,,
\ee
where $\widehat{m}_{1,2}$ measure the mass of each hemisphere jet on a given final state:
\be
\widehat{m}_{1,2} \ket{X_{\text{hemi 1,2}}} = m_{1,2} \ket{X_{\text{hemi 1,2}}} \,.
\ee
In SCET an $N$-jet final state is described by an operator, $O_N$, with N-collinear sectors coupled to soft modes which are exchanged between them. For the dijet invariant mass, the SCET matrix element is
\be
\langle 0 \lvert \, O_2^{\dagger} \, \R(m_1,m_2) \, O_2 \, \rvert 0 \rangle \,.
\ee
In order to factorize the cross section in SCET we have to carry out steps 2 through 4 in the above list. We must factorize the 2-jet operator and the restriction operator into soft and collinear pieces.  

The BPS field redefinition performs soft-collinear decoupling of the operator, factorizing $O_2$ schematically in to
\be
O_2 \to [ O_{c_1}] [O_{c_2}] [ O_s] \,,
\ee
where the two collinear sectors $c_{1,2}$ do not couple to each other or the soft operator $O_s$ to leading power in $\lambda$. This satisfies step 2 for any 2-jet observable of interest.

Factorization of $\R$ divides the operator into separate measurements in the collinear and soft sectors.  At a formal level, this factorizes the operator:
\be \label{eq:Rfact}
\R = \R_s \otimes \sum_i \R_c^i \,,
\ee
where the convolution comes from the observable dependence in each operator. The soft restriction operator $\R_s$ acts only on soft modes, while the collinear restriction operator $\R_c^i$ acts only on the collinear modes in sector $i$.  OPS factorization tests whether this holds by examining the phase space constraints in each sector.  For the jet mass, this is straightforward: from \eq{Mjet}, the soft and collinear particles in the jet linearly contribute to the mass and the jet boundaries enforced by the measurement are determined by the jet directions.  This satisfies step 3 in the above list.  For jet algorithms, OPS factorization requires showing that the constraints that specify which soft and collinear particles are put in the jet are independent of the other sectors.  For jet substructure, it requires showing that the kinematic cuts placed on the jet constituents separate into cuts on the soft and collinear phase space.  For example, cuts that remove wide angle soft radiation in the jet must be independent of the collinear phase space.
 
Formally, factorization of the restriction operator requires providing operator definitions for $\widehat{m}^{c,s}_{1,2}$ in step 4.  For the jet mass (equivalent to thrust $\tau$ in the small mass limit), these operators were constructed explicitly from the energy-momentum tensor in \cite{Bauer:2008dt}.  This, coupled with the arguments of soft-collinear factorization, shows
\be \label{eq:mfact}
\widehat{m}_{1,2} = \widehat{m}_{1,2}^c + \widehat{m}_{1,2}^s \,,
\ee
from which \eq{Rfact} follows.  Factorization of the restriction operator allows us to write the cross section in the form
\begin{align} \label{eq:sigmafact}
\sigma (m_1, &m_2) = H_2  \int {\rm d}\, m_1 {\rm d} \, m_2 \, \delta(m_1-m_1^c-m_1^s) \\ 
& \delta(m_2-m_2^c-m_2^s)J_n (m_1^c) \, J_{\bn} (m_2^c) S_2 (m_1^s, m_2^s) \,. \nn
\end{align}
The jet and soft functions are forward scattering matrix elements of soft or collinear fields with restriction operators that act only on the phase space of that sector.  We will explore OPS factorization with several jet algorithm and substructure examples in the remainder of this paper.

\section{A Test for Soft-Collinear Factorization}
\label{sec:tests}

In this section we analyze a necessary condition for factorization relevant to a wide class of jet observables where the soft and collinear dynamics of \SCETI provide the relevant description. We focus on a key step in the proof of factorization as outlined in \subsec{factoverview}, OPS factorization. This is powerful at detecting when factorization fails, and it is sensitively dependent on the precise definition of the jet observable.  It is also very useful in characterizing the behavior of algorithms or substructure.
  
The jet and soft functions give the contribution of soft and collinear modes to the cross section.  These functions implement the phase space constraints from the observable, which can be expressed as a restriction operator $\R$ that acts on the final state.  OPS factorization splits the restriction operator into separate pieces that only act on modes in a given sector, as in \eq{Rfact}.  Proving OPS factorization requires showing the following:
\begin{list}{\labelitemi}{\leftmargin=1.5em}
\item To leading order in the power counting parameter $\lambda$, the phase space constraints on soft and collinear modes must be independent of each other to all orders in perturbation theory.
\end{list}
We can test this by carrying out a simple power counting analysis on the algorithm or observable, and generally find that it can be very constraining.  

Note that the soft modes are not completely ignorant of the collinear modes.  The soft Wilson line knows about the direction of the associated collinear sector, and the measurement can add dependence on the total energy in each collinear sector.  However, the phase space constraints in the soft sector must be independent of individual collinear splittings or momenta.  Similarly, the collinear phase space constraints must be independent of individual soft momenta.   If the phase space constraints are not independent, then the measurement operator cannot factorize, as they do in the example in \eq{mfact}, since the action of $\R_s$ or $\R_c$ would depend on both collinear and soft states.

The factorization in \eq{Rfact} must hold at all orders in $\alpha_s$; it is not sufficient to consider cases at lowest order in perturbation theory.  However, we will find studying fixed order configurations can be very useful and can often be generalized to all-orders arguments.  For instance, since recombination-style jet algorithms deal with pairs of particles, often simple $\ca{O}(\alpha_s^2)$ configurations are sufficient to make broader statements about the behavior and factorizability of an algorithm.

Because jets cover a wide kinematic range, factorization constraints will depend on the types of cuts placed on the jet.  The additional scales present in jet substructure algorithms can require factorization that goes beyond the simple soft-collinear factorization that we discuss here.  In any case, we must show that step 1 in the list from \sec{fact} still holds, which now may require additional modes.  An example relevant for jet substructure is the case where two jets (or subjets) come close together \cite{Bauer:2011uc}.  This factorization is described by \SCETp, an extension of \SCETI that includes an additional mode to describe soft radiation into the nearby jets.  Another possibility is that a different description of soft and collinear modes is needed, such as \SCETII, where the components of soft momentum scale as $\lambda$ instead of $\lambda^2$.  However, soft and collinear modes are still at the heart of jet physics problems, and OPS factorization is still a key part of factorization in these cases.  We discuss alternative forms of factorization in more detail in \sec{caveats}.

\subsection{The Structure of Jets in SCET}
\label{subsec:general}

In this section we highlight some of the generic constraints imposed by OPS factorization on jet algorithms and substructure methods.  We use the structure of the jet implied by power counting of the soft and collinear modes in SCET.  

\begin{figure}[h]
\centerline{\scalebox{0.45}{\includegraphics{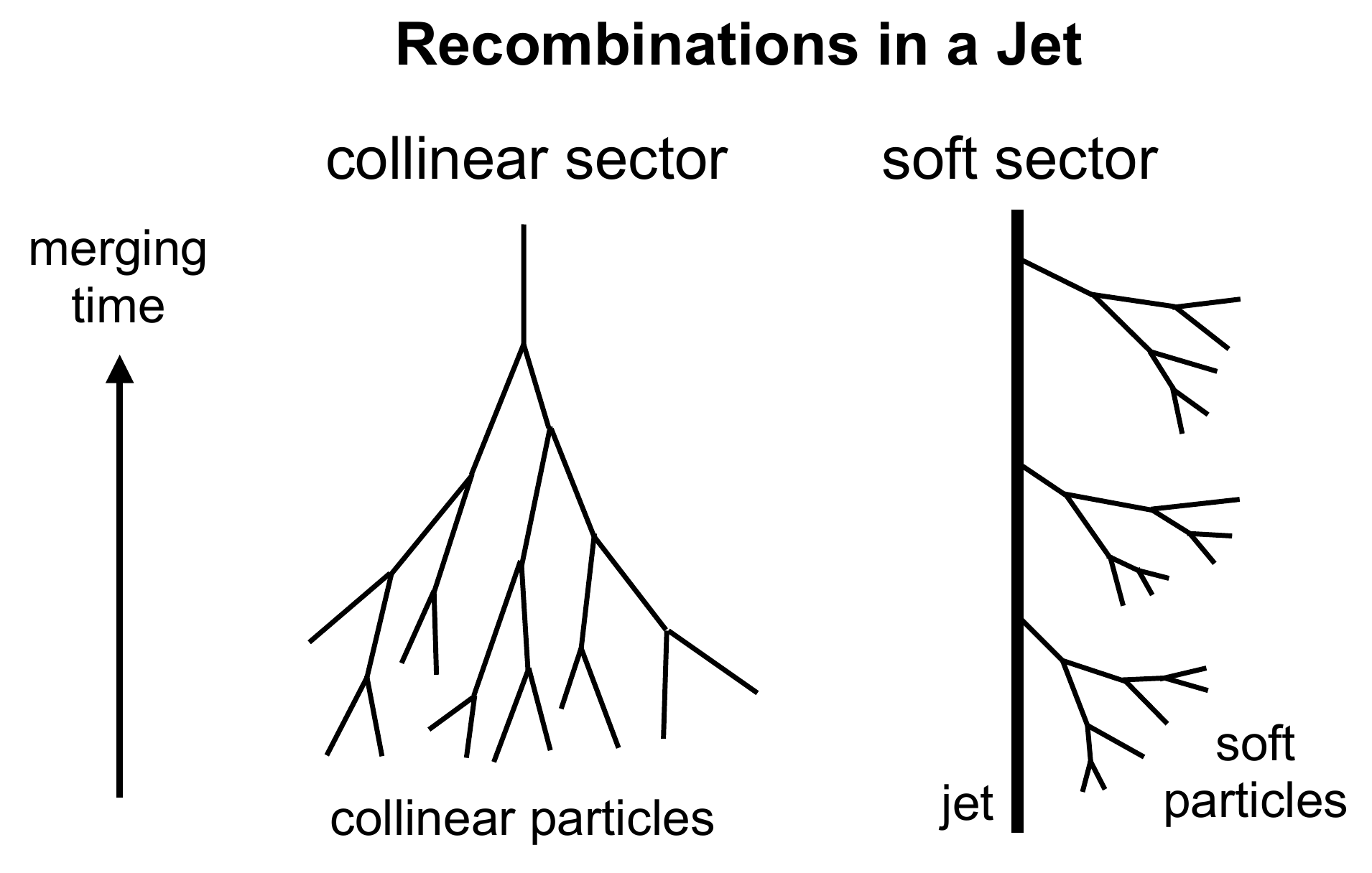}}}
\caption{Recombination structure of the soft and collinear particles in a jet.  Because the energy and direction of a collinear particle is unchanged after merging with a soft particle, soft-collinear recombinations are represented in the soft tree.  And, since soft modes can only resolve the jet direction and not individual collinear particles, the collinear mode in these recombinations is represented by a thick line for the jet.} \label{fig:algmerging}
\end{figure}

Consider a jet found by a recombination algorithm, which builds jets through $2\to1$ merging of particles.  The recombinations that make the jet in SCET consist of three different kinds of merging of soft and collinear modes:
\be
c, c \to c \,, \qquad c, s \to c \,, \qquad s, s \to s \,.
\ee
It is useful to depict the sequence of mergings as a tree. This illustrates the action of the algorithm, which generates the phase space constraints of the jet and should not be mistaken with the jet evolution. In QCD each jet has its own tree.  However, with soft and collinear modes we have separate trees for each type of mode. Each jet in the event is therefore represented by two recombination trees.  Collinear-collinear recombinations are confined to the tree of collinear particles, and soft-soft recombinations are confined to the tree of soft particles.  Soft-collinear recombinations are special; recall \eq{softindep}, which shows that the collinear energy and direction are unchanged up to power corrections.  Therefore, the $c, s \to c$ merging is represented in the soft tree.  Since the soft mode can only resolve the jet direction, and not the direction of the specific collinear mode, the soft particle only sees that it is recombining with ``the jet'', represented by a Wilson line in the EFT.  This is equivalent to the statement in \eq{nsangle}, that $\Delta R_{cs} = \Delta R_{ns} +\ca{O}(\lambda^2)$.  The structure of the trees of soft and collinear modes is shown in \fig{algmerging}.

Given the recombination structure of a jet in SCET, it is easy to see that OPS factorization implies that the cross section cannot know about the relative merging ordering in each tree. The soft function only knows about the recombinations in the soft tree, and each jet function only knows about the recombinations in a collinear tree.  The relative order of recombinations in each tree is not available to any function in a factorized cross section.  This means that, for example, an observable that depends on the \textit{last} recombination in the jet (the $2\to1$ recombination that forms the jet) cannot be factorized, since we cannot even know the type of recombination ($c,c \to c$ or $c,s \to c$).  This will have obvious consequences for jet substructure methods that depend on declustering the jet.

A natural object that jet substructure methods deal with are subjets, which are clusters of particles within the jet. Collinear subjets contain both collinear and soft particles, and have the momentum scaling of collinear particles. Soft subjets contain only soft particles, and so contain contributions only from the soft tree; their momentum scales like soft particles. This terminology is useful in the following sections.

\subsection{Examples from Jet Algorithms}
\label{subsec:jetalgs}

As a warm-up to jet substructure, we apply the ideas of soft-collinear factorization to jet algorithms.  They provide a familiar and simple example, and the power-counting analysis of OPS factorization gives insight into the leading behavior of the algorithms.  Our primary examples are the $\kt$, Cambridge/Aachen, and anti-$\kt$ algorithms, all of which factorize.  SCET factorization constraints for these algorithms were explored in \cite{Ellis:2010rwa}.  We also consider the JADE algorithm, which provides a useful example of an algorithm that does not factorize.

All of the algorithms we consider are recombination algorithms.  These algorithms use two metrics:
\begin{align}
\rho_{ij} &\textrm{: distance between }i\textrm{ and }j \,, \nn \\
\rho_i &\textrm{: single particle metric for }i \,.
\end{align}
The algorithm operates by finding the smallest of all $\rho_{ij}$ and $\rho_i$.  If the smallest is a pairwise metric $\rho_{ij}$, particles $i$ and $j$ are merged by adding their four-momentum.  If the smallest is a single particle metric $\rho_i$, then particle $i$ is promoted to a candidate jet and removed from the set of particles.  This process is repeated until all particles have been combined into candidate jets, and only candidate jets with energy or $p_T$ greater than a cutoff are counted as final state jets.

Since the jet is made up of collinear and soft particles, there are multiple kinds of pairwise and single particle metrics.  The pairwise metrics come from collinear-collinear, collinear-soft, and soft-soft pairs:
\be
\rho_{cc} \,, \,\, \rho_{cs} \,, \,\, \rho_{ss} \,.
\ee
There are also collinear and soft single particle metrics:
\be
\rho_c \,, \,\, \rho_s \,.
\ee
The scaling behavior of these metrics determine the behavior and factorization properties of the jet algorithm.

\subsubsection{The $\kt$ Class of Algorithms}
\label{subsec:kTclass}

The most common recombination algorithms are the $\kt$ class of algorithms.  They are parameterized by a number $\alpha$, where $\alpha = 1$ is the $\kt$ algorithm, $\alpha = 0$ is Cambridge/Aachen, and $\alpha = -1$ is the anti-$\kt$ algorithm.  The metrics for these algorithms are
\begin{align} \label{eq:kTclassmetrics}
\rho_{ij} &= \min\left( p_{Ti}^{\alpha}, p_{Tj}^{\alpha} \right) \Delta R_{ij} \,, \nn \\
\rho_i &= p_{Ti}^{\alpha} \, R \,.
\end{align}
The $\kt$, C/A, and anti-$\kt$ algorithms produce very different behavior in the structure of the jet.
\renewcommand{\arraystretch}{2.5}
\renewcommand{\tabcolsep}{9pt}
\begin{table*}[htdp]\normalsize
\caption{\normalsize{Metric scaling for the $\kt$ class of jet algorithms.}}
\begin{center}
\begin{tabular}{|c|c|c|c|}
\hline
metrics & $\kt$ $\quad \alpha = 1$ & C/A $\quad \alpha = 0$ & anti-$\kt$ $\quad \alpha = -1$ \\
\hline
\hline
$\rho_{ci\, cj}$ & $\min\left(p_{Tci}, p_{Tcj} \right) \Delta R_{ci\, cj} \sim \ca{O}(\lambda) $ & $\Delta R_{ci\, cj} \sim \ca{O} (\lambda)$ & $ \min\left( \dfrac{1}{p_{Tci}} , \dfrac{1}{p_{Tcj}} \right) \Delta R_{ci\, cj} \sim \ca{O}(\lambda) $ \\ 
[1ex]\hline
$\rho_{cs}$ & $p_{Ts} \Delta R_{ns} \sim \ca{O}(\lambda^2)$ & $\Delta R_{ns} \sim \ca{O}(1)$ & $\dfrac{1}{p_{Tc}} \Delta R_{ns} \sim \ca{O}(1)$ \\
[1ex]\hline
$\rho_{si\, sj}$ & $\min\left(p_{Tsi}, p_{Tsj}\right) \Delta R_{si\, sj} \sim \ca{O}(\lambda^2)$ & $\Delta R_{si\, sj} \sim \ca{O}(1)$ & $\min\left(\dfrac{1}{p_{Tsi}}, \dfrac{1}{p_{Tsj}} \right) \Delta R_{si\, sj} \sim \ca{O}(\lambda^{-2})$ \\
[1ex]\hline
$\rho_c$ & $p_{Tc} R \sim \ca{O}(R)$ & $R$ & $\dfrac{1}{p_{Tc}} R \sim \ca{O}(R)$ \\
[1ex]\hline
$\rho_s$ & $p_{Ts} R \sim \ca{O}(\lambda^2 R)$ & $R$ & $\dfrac{1}{p_{Ts}} R \sim \ca{O}(\lambda^{-2} R)$ \\
[1ex]\hline
\end{tabular}
\end{center}
\label{table:algscaling}
\end{table*}%

To understand the factorization properties of the algorithms, we determine the characteristic scaling of $\rho_{cc}$, $\rho_{cs}$, and $\rho_{ss}$ for central jets.  These are given in Table~\ref{table:algscaling}, and in \fig{AlgorithmOrdering:metrics} we show the relative ranking of the pairwise metrics for each algorithm. As noted in \subsec{SCEToverview}, it is possible for the pairwise metric to have a scaling smaller than the characteristic scaling shown while contributing to the jet at leading power in $\lambda$. We consider this where relevant in the following discussion.

OPS factorization requires that the comparison of the metrics needed to run the algorithm decouple soft and collinear phase space constraints.  For the $\kt$ and C/A algorithms, this is straightforward.  First, note that the pairwise metrics $\rho_{ss}$ and $\rho_{cs}$ depend only on soft momenta and the jet direction $\vec{n}$, through $\Delta R_{ns}$.  Although naively the phase space constraints from comparing $\rho_{cc}$ with $\rho_{cs}$ or $\rho_{ss}$ would ruin factorization (because we are making a comparison whose value depends on both collinear and soft momenta), this ordering is in fact irrelevant.  From \eq{softindep}, the energy and direction of a collinear particle is the same at leading order before and after a soft-collinear recombination.  This implies that collinear-collinear recombinations and soft-collinear or soft-soft recombinations occur independently of each other, and which particles are merged into the final state jet does not depend on the relative ordering.

As observed in \subsec{general}, the factorized cross section cannot know about this ordering.  This gives a simple, general constraint on a factorizable jet algorithm: at leading order in the power counting, the set of particles included in a jet cannot depend on the relative ordering of $\rho_{cc}$ and $\rho_{cs}$ or $\rho_{ss}$.
\begin{figure}[h]
\centerline{\scalebox{0.3}{\includegraphics{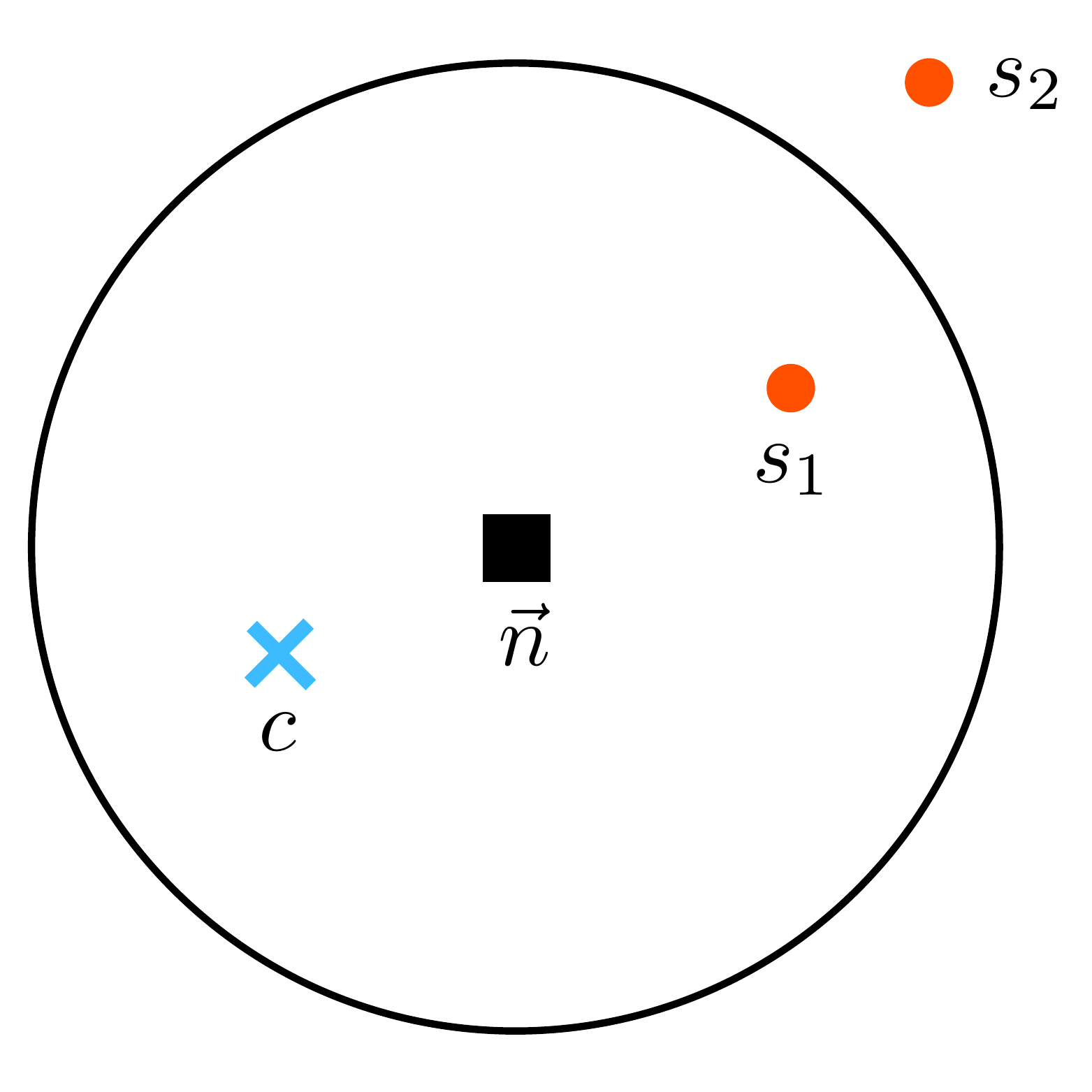}}}
\caption{A jet of size $R$ in the $\vec{n}$ direction, shown schematically in the $y-\phi$ plane.  A collinear particle, $c$, and two soft particles, $s_1$ and $s_2$, are shown.
} \label{fig:cssConfig}
\end{figure}

OPS factorization for the anti-$\kt$ algorithm is slightly more complicated since $\rho_{cs}$ depends on both soft and collinear momenta.  As with the $\kt$ and C/A algorithms, the ordering of $\rho_{cc}$ and $\rho_{cs}$ or $\rho_{ss}$ is irrelevant.  However, the comparison between $\rho_{cs}$ and $\rho_{ss}$ for the anti-$\kt$ algorithm would naively spoil factorization.  We show that factorization holds by considering a simple configuration shown in \fig{cssConfig}, which can be easily generalized.  Consider a collinear particle $c$ and soft particles $s_1$ and $s_2$, where
\be
\Delta R_{ns_1} < R < \Delta R_{ns_2} \,.
\ee
If $\rho_{c\, s_1} < \rho_{s_1 \, s_2}$, then the $c-s_1$ recombination occurs first and the jet does not include $s_2$.  But if $\rho_{s_1\, s_2} < \rho_{c\, s_1}$, then the $s_1-s_2$ recombination puts both soft modes either in or out of the jet.  However, the region of phase space for which the $s_1 - s_2$ recombination occurs first is power suppressed.  The condition on the soft-soft angle that this occurs is
\be
\Delta R_{s_1\, s_2} < \frac{\max\left(p_{T{s_1}} , p_{T{s_2}} \right)}{p_{Tc}} \Delta R_{n \,{s_1}} \lesssim \ca{O}(\lambda^2 R) \,.
\ee
Since the configuration we are considering has $\Delta R_{n\, s_2} > R$, this means that only region of phase space where the soft-soft recombination occurs first is when the two soft modes are located at the jet boundary, separated by an angle of $\ca{O}(\lambda^2 R)$.  This region of phase space is power suppressed since the total area of this region is parametrically smaller than the jet area.  This means these configurations do not contribute to the rate at leading power, and can be ignored in the same approximation.  Therefore, the OPS factorization constraints on the pairwise metrics are satisfied.

Finally, for all the algorithms one may be concerned that comparisons between the single and pairwise metrics could spoil factorization.  However, for energetic jets (jets that contain collinear particles), the single particle metric only serves to enforce the constraint that two particles can be recombined only if their angular separation is less than $R$.  For soft jets, since there are no collinear modes the phase space only depends on soft momenta and trivially factorizes.

We have seen, using a simple power counting analysis with soft and collinear modes, that the $\kt$, C/A, and anti-$\kt$ algorithms have OPS factorization.  These algorithms can be formally factorized in SCET by providing an operator definition for the phase space restrictions that the algorithm implements.  We turn now to the dominant behavior of these algorithms.

\subsubsection{Characteristic Behavior of the $\kt$ Class of Algorithms}
\label{subsec:algordering}

For each algorithm the ordering of the parametric scaling of the metrics indicate the order in which recombinations tend to occur. This gives the characteristic behavior of the algorithm.  This behavior is well known, but it is instructive to see it arise from a simple power counting analysis, and the same arguments will be useful for jet substructure.  We need only consider the pairwise metrics, since the single particle metrics simply set the size of the jet.  We also only deal with energetic jets, as soft jets have a simpler behavior but are usually removed by $p_T$ cuts.

For the $\kt$ algorithm, $\rho_{ss} \sim \rho_{cs} \ll \rho_{cc}$.  This means that the first step in the algorithm is to merge soft particles with soft and collinear particles.  Eventually only collinear particles will remain, and the next step of the algorithm merges them to form jets.  Since soft particles are merged first, the jet boundary is determined by the local clusters of soft radiation at the edge of the jet.  This gives rise to the ``vacuuming'' effect of the $\kt$ algorithm, where the boundary of the jet is amorphous and tends to include soft radiation farther than $R$ from the jet axis.  

For the C/A algorithm, $\rho_{cc} \ll \rho_{ss} \sim \rho_{cs}$.  In this case collinear particles will be merged first, and they will merge into a single collinear object that has the jet energy and direction (up to power corrections).  Then soft particles will merge among themselves and with the jet.  Just like the $\kt$ algorithm, the shape of the jet boundary is determined by the local soft-soft recombinations.  However, since the metric of the algorithm weights only by angle and not by $p_T$ (as with $\kt$), the amount of vacuuming with the C/A algorithm is less than the $\kt$ algorithm, since soft particles at the periphery of the jet are less frequently merged into the jet.

For the anti-$\kt$ algorithm, $\rho_{cc} \ll \rho_{cs} \ll \rho_{ss}$.  As with the C/A algorithm, collinear particles are merged to form a single collinear mode (the jet) first.  Since $\rho_{cs} \ll \rho_{ss}$, the next step in the anti-$\kt$ algorithm is to merge all soft particles within a radius $R$ of the jet axis.  At this point nothing else will be merged into the jet, since all remaining soft particles are too far from the jet axis.  Soft-soft recombinations are only relevant for pairs of soft particles whose separation is parametrically smaller than the scaling would indicate, or for jets consisting of only soft particles.  This means that the anti-$\kt$ jets will be very circular, as is well-known.

We can compare the behavior of the algorithms implied by the parametric scaling to the behavior in Monte Carlo simulation.  We use the Boost 2010 event samples\footnote{These events are publicly available from two repositories, hosted by {\href{http://www.lpthe.jussieu.fr/~salam/projects/boost2010-events/}{\color{blue}{Gavin Salam}}} and {\href{http://tev4.phys.washington.edu/TeraScale/boost2010/}{\color{blue}{The University of Washington}}}.} here and for the Monte Carlo study of pruning behavior in \sec{jetsub} \cite{Abdesselam:2010pt}.  The particular sample we use is simulated QCD events with hard partons between 500 and 600 GeV, generated with Pythia v6.421 using the DW tune.  We implement the standard analysis cuts suggested in the Boost 2010 report, keeping only visible particles (except muons) with $|\eta| < 5.0$.  Only the two hardest jets with $p_T > 200$ GeV are used.  We use FastJet 3.0$\beta$1 to cluster the jets and look at the substructure \cite{Cacciari:2005hq}.

\begin{figure*}[th]
\subfigure[]{\centerline{\scalebox{0.33}{\includegraphics{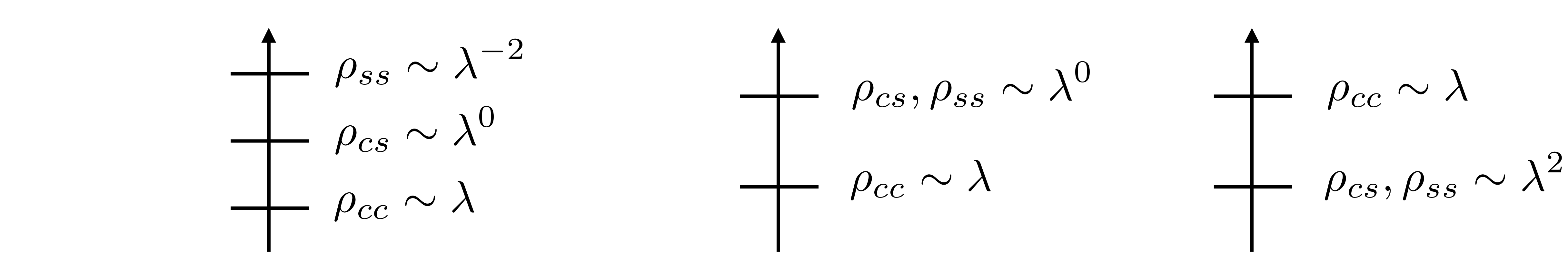}}}\label{fig:AlgorithmOrdering:metrics}}
\subfigure[]{\centerline{\scalebox{0.39}{\includegraphics{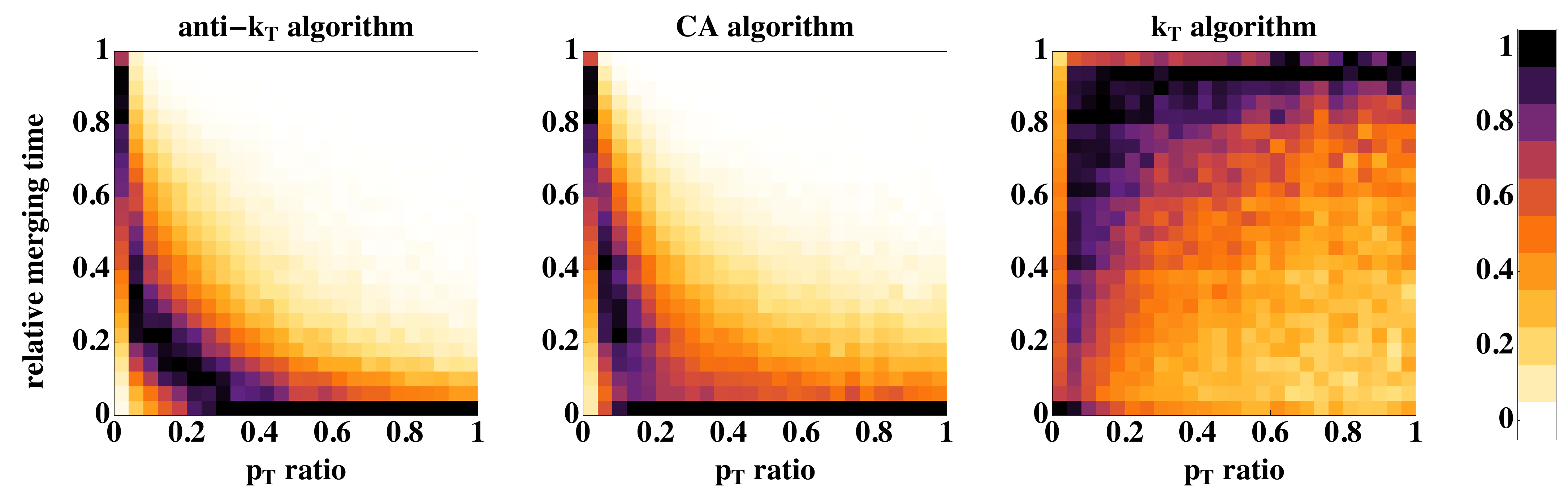}}}\label{fig:AlgorithmOrdering:plot}}
\caption{(a) Relative ordering of the pairwise metrics for the anti-$\kt$, C/A, and $\kt$ algorithms at leading power. (b) Relative merging times vs.~particle $p_T$ ratio for anti-$\kt$, C/A, and $\kt$ jets, using jets from the Boost 2010 sample described in the text.  For each algorithm, the jets are found with using $R=1.0$ and have $p_T > 200$ GeV and $m/p_T < 0.25$.  The density in each column of bins for fixed scaled particle $p_T$ is separately normalized.  The relative merging times agree well with the qualitative ordering of the pairwise metrics.} \label{fig:AlgorithmOrdering}
\end{figure*}

Define the relative merging time for particle $i$ in a jet to be
\be
\textrm{relative merging time } = \frac{n^{i}_{\rm step}}{n_{\rm step}^J} \,,
\ee
where $n^{J}_{\rm step}$ is the number of recombination steps needed to make the jet and $n^{i}_{\rm step}$ is the step number at which particle $i$ was first merged with another particle in the jet.  The first particles to merge have relative merging time 0, while a particle that is not recombined until the end of the algorithm has relative merging time 1.  In \fig{AlgorithmOrdering:plot}, we cluster jets using $R = 1.0$ with each algorithm and select those with $m/p_T < 0.25$.  We then plot the merging time versus the ratio of the particle's $p_T$ to the highest $p_T$ particle in the jet for the three algorithms.  The figure shows the density of particles in these two variables, and because there are many more particles with small $p_T$ than large $p_T$ we normalize the density separately in each bin of $p_T$ ratio.

From the figure, we can see that the parametric ordering of recombinations of each algorithm matches well to the Monte Carlo events.  The Monte Carlo events have the same ordering of merging time for soft and collinear particles predicted by power counting, and the particles are fairly tightly clustered around their parametric merging time.  Recall that this analysis is based on the parametric scaling of the metric, and leading order contributions away from this scaling can give corrections to the behavior shown here.  While the parametric scaling is only the dominant action of the algorithm, it is encouraging that the picture of soft and collinear modes captures the behavior of the algorithm well.

\subsubsection{The JADE Algorithm}
\label{subsec:JADE}

Finally we consider the original recombination algorithm, JADE \cite{Bartel:1986ua,Bethke:1988zc}.  JADE is an \textit{exclusive} algorithm, meaning the algorithm operates with a pairwise metric $y_{ij}$ and a cut parameter $y_{\rm cut}$ (instead of a single particle metric).  When the smallest $y_{ij}$ is greater than $y_{\rm cut}$, the algorithm stops and all remaining particles are promoted to jets.

The pairwise metric for JADE is the invariant mass,
\be \label{eq:JADEmetric}
y_{ij} = \frac{1}{Q^2} (p_i + p_j)^2 \,.
\ee
Although JADE is a well-defined, infrared safe jet algorithm, it was shown by explicit calculation at $\ca{O}(\alpha_s^2)$ that the leading logarithms in the dijet cross section, $\alpha^n \ln^{2n} y_{\rm cut}$, do not exponentiate \cite{Brown:1990nm}.  This spoils the perturbative expansion and prevents accurate theoretical predictions from being made. This calculation is an explicit demonstration that JADE does not factorize.  Furthermore, it was shown that JADE does not satisfy a necessary and sufficient condition for exponentiation to next-to-leading logarithmic accuracy, recursive infrared and collinear (rIRC) safety \cite{Banfi:2004yd}. The failure of factorization of JADE is easy to see using a power-counting analysis of OPS factorization, which we now show. 

The soft and collinear modes in the JADE metric scale at leading power as
\begin{align} \label{eq:JADEscaling}
&y_{ci\, cj} = \frac{(p_{ci} + p_{cj})^2}{Q^2} \sim \ca{O}(\lambda^2) \,, \nn \\
& y_{c s} = \frac{(p_c + p_s)^2}{Q^2} = \frac{(p_c^2 + p_c^- p_s^+)}{Q^2} [1+\ca{O}(
\lambda^2)] \sim \ca{O}(\lambda^2) \,, \nn \\
& y_{si\, sj} = \frac{(p_{si} + p_{sj})^2}{Q^2} \sim \ca{O}(\lambda^4) \,.
\end{align}
In order for collinear particles in a jet to be recombined while maintaining collimated jets, $y_{\rm cut}$ should be $\ca{O}(\lambda^2)$. The typical action of the algorithm is to combine soft particles first, followed by soft-collinear and collinear-collinear recombinations.  In running the algorithm we will compare the soft-collinear metric $y_{cs}$ to $y_{\rm cut}$.  Since $y_{cs}$ depends on the momenta of both soft and collinear final state particles, this comparison cannot be factorized into separate phase space constraints on the soft and collinear sectors.  Since the action of the algorithm depends on the value of the soft-collinear metric $y_{cs}$, the algorithm cannot factorize.

\begin{figure}[t]
\centering
\includegraphics[scale=0.33]{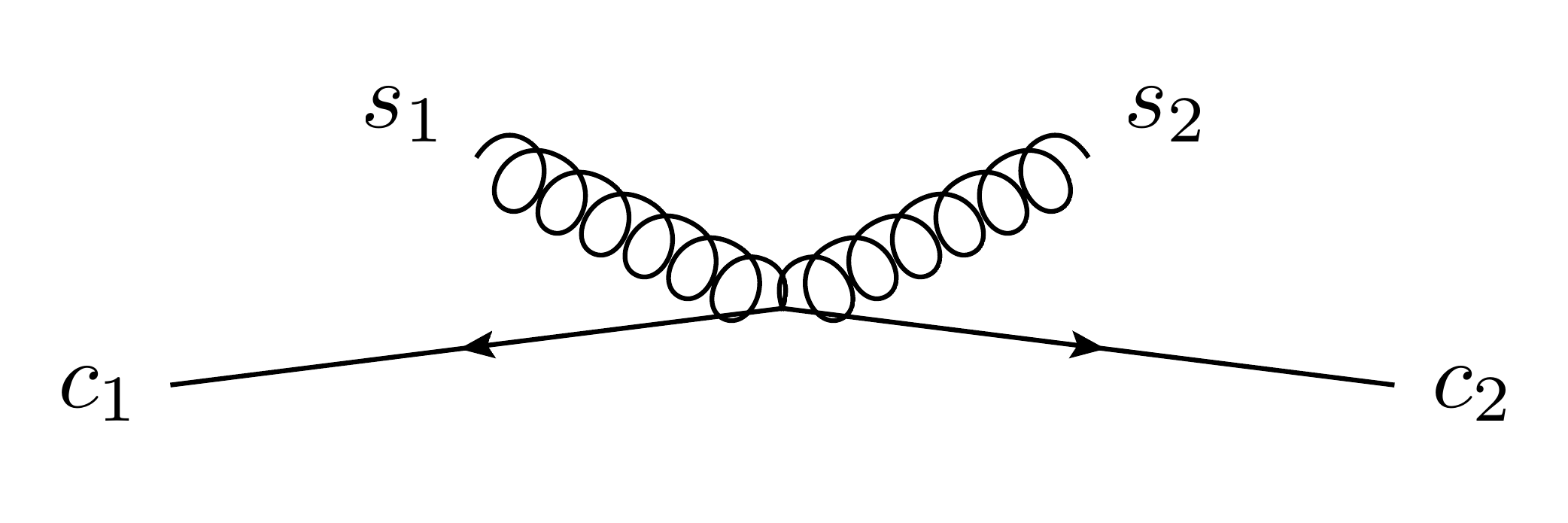}
\vspace{-0.5ex}
\caption{A simple configuration that shows the failure of soft-collinear factorization of JADE \cite{Brown:1990nm}.  The algorithm combines the soft gluons $s_1$ and $s_2$ first.  There is a region of phase space that contributes to the leading log 3-jet rate where the soft gluon pair does not get recombined with either collinear quark.}
\label{fig:JADEconfig}
\end{figure}

The standard example, from Brown and Stirling \cite{Brown:1990nm}, that demonstrates the problem with the JADE algorithm is shown in \fig{JADEconfig}. When analyzed using power counting in SCET, it illustrates the failure of OPS factorization discussed above at fixed order in perturbation theory. This 4-parton configuration has two energetic quarks and two soft gluons with
\be
y_{s_1 \, s_2} < \{ y_{c_1\, s_1} , \, y_{c_2\, s_2} \} < y_{\rm cut} \,.
\ee
JADE will recombine the soft gluons first, and after this recombination there exists a region of phase space where
\be \label{eq:JADEproblem}
\{ y_{c_1 \, s_{12}} ,\, y_{c_2\, s_{12}} \} > y_{\rm cut} \,,
\ee
making the event a 3-jet event. The comparison in \eq{JADEproblem} is precisely what leads to the failure of factorization of JADE. We note that this problem  already arises at $\ca{O}(\alpha_s)$. The failure of factorization in JADE is particularly severe since it spoils resummation of even the leading logarithmic series. With only one of the soft gluons in \fig{JADEconfig} (at $\ca{O}(\alpha_s)$) the event is a 2-jet event, but with two soft gluons (at $\ca{O}(\alpha_s^2)$) this region of phase space can contribute to the 3-jet rate at leading logarithmic accuracy. 

\section{Factorization for Jet Substructure}
\label{sec:jetsub}

We now consider factorization constraints for jet substructure.  The general constraints on merging order in \subsec{general} apply to substructure algorithms, and the lessons from the jet algorithm examples will be useful.  We consider four different substructure algorithms: the mass-drop filter algorithm, pruning, trimming, and $N$-subjettiness.  The MD-F algorithm uses two generic procedures, declustering and filtering, common to many substructure algorithms.  The pruning and trimming algorithms are generic grooming procedures, designed to remove soft radiation far from energetic clusters.  $N$-subjettiness is an example of a jet shape useful for substructure, for which the factorization analysis is simpler.

Traditional substructure algorithms use broad approaches to improve the discrimination over QCD jets.  Many algorithms place kinematic cuts on the hard splitting of the boosted object to identify subjets.  For example, the Johns Hopkins top-tagger reduces the QCD background by placing cuts to identify the top and $W$.  It requires a hard subjet to have a mass near $m_{\text{top}}$, a daughter subjet have a mass near $m_W$, and places a cut on the helicity angle of the subjets of the $W$ \cite{Kaplan:2008ie}.  Substructure algorithms can groom the jet by removing soft, wide-angle radiation that is characteristic of the underlying event and pileup.  Finally, they also discriminate QCD and non-QCD jets based on the radiation pattern in the jet, identifying characteristics of a particular decay based on properties such as color.

Recently, jet shapes have been shown to be effective jet substructure tools \cite{Thaler:2010tr,Jankowiak:2011qa,Thaler:2011gf}.  Jet shapes define an observable from a projection of the momenta in a jet, making a direct measurement of the jet without manipulating the jet's constituents.  This makes the factorization analysis simple for basic shapes, but subtleties remain for more complex shapes.

Factorization constrains the scaling of parameters in all of the traditional substructure algorithms, and we find that declustering and filtering do not factorize.  However, simple modifications derived from a power counting analysis of the algorithm can be made to allow for factorization.  We also find that with the proper scaling for the parameters of pruning and trimming, jet shapes (such as the jet mass) pass the test for observable factorization.  More exclusive observables such as subjet masses do not factorize without additional kinematic restrictions enforcing the subjets be well-separated. This allows them to be treated as separate collinear sectors and relaxes the factorization constraints.

We find it convenient to first discuss pruning and trimming, as the power counting analysis is straightforward to apply.  This analysis also produces a simple picture of the phase space remaining after pruning.  We compare the predictions from power counting to Monte Carlo simulation and find good agreement.  Next, we discuss the MD-F algorithm.  After identifying the problems from factorization with declustering and filtering, we explore modifications that can allow for factorization.  We compare the modified MD-F algorithm to the original using a simple Monte Carlo analysis.  Because many substructure algorithms are based on the declustering and filtering steps in MD-F, the issues we discuss for MD-F are more widely relevant.  Finally, we discuss generic jet shapes and their factorization constraints, specializing to the case of $N$-subjettiness.  In general, jet shapes are more theoretically tractable and more amenable to calculation, and many jet shape calculations for the LHC already exist.

\subsection{Pruning}
\label{subsec:pruning}

The pruning algorithm is intended to improve the mass resolution of boosted heavy particles decaying to single jets by selectively removing isolated soft radiation in jets \cite{Ellis:2009su,Ellis:2009me}.  Pruning uses the fact that isolated soft radiation can contribute significantly to poor mass resolution in jets, and that recombination algorithms naturally identify this radiation.  In pruning, jets are reclustered and kinematic cuts are placed on each reclustering step; soft particles at wide angles to energetic ones are removed from the jet.  This reduces the QCD background to heavy particles and allows the substructure of the jet to be used to reconstruct the decay of heavy particles.

Pruning removes particles through a secondary clustering procedure.  Starting with jets found with large $R$, the pruning procedure is:
\begin{list}{\labelitemi}{\leftmargin=1.5em}
\item[1.] Recluster the jets with a recombination algorithm.  At each recombination $i,j\to i+j$, test if
\be
z_{ij} \equiv \frac{\min(p_{T,i},p_{T,j})}{p_{T,i+j}} < z_{\rm cut} \quad \textrm{and} \quad \Delta R_{ij} > D_{\rm cut} \, .
\ee
\item[2.] If both of these conditions are met, then discard the softer of $i,j$ and continue reclustering without any merging.  Otherwise, recombine the pair and continue.  The jet that is formed is the new, pruned jet.
\end{list}
The reclustering algorithm is defined without a promotion metric, so that every particle in the jet ends up being reclustered or pruned (discarded).  The dimensionless parameters $z_{\rm cut}$ and $D_{\rm cut}$ are inputs to the algorithm.  In the pruning definition, $D_{\rm cut} = m_J/p_{TJ}$ was suggested, as was $z_{\rm cut} \sim 0.10 - 0.15$.

Pruning is slightly different from other substructure methods since it reclusters the jet and applies kinematic cuts at each stage of reclustering, instead of first finding subjets and then implementing kinematic cuts.  This makes the factorization considerations more straightforward, since we can consider the effects of the pruning cuts at each clustering step.

We first derive a general constraint on substructure algorithms:
\begin{list}{\labelitemi}{\leftmargin=1.5em}
\item A collinear subjet cannot be removed from the jet unless the entire collinear sector (jet or subjet) is removed.
\end{list}
A generic collinear subjet, represented by a branch in the collinear tree in \fig{algmerging}, has been merged with soft particles by the algorithm.  As discussed in \subsec{general}, we cannot know \textit{which} soft particles have been merged into that collinear subjet: soft particles cannot resolve individual collinear particles.  Therefore, when we remove the subjet, the affect on the soft function is ambiguous.  Since the phase space constraint that removes the collinear subjet is implemented in the jet function, OPS factorization requires that the soft sector cannot know about this cut.  Therefore, removing collinear subjets will break factorization.

The only way to evade this constraint is if an entire collinear sector is removed.  If an entire collinear sector is removed, then all soft particles merged into the sector will be removed.  This does not mean that the entire jet has to be discarded, however.  If we implement phase space cuts so that a collinear subjet is well separated from all other collinear particles in the jet, then we can treat that subjet as its own collinear sector.  By well-separated, we mean that the subjets are collimated with respect to their separation, e.g., $m_{j_1 j_2} \gg \{m_{j_1}, m_{j_2}\}$.  In general, though, it is not physically sensible to remove an entire collinear sector.  Collinear radiation typically comes from energetic particles in a jet's evolution, and is a part of any useful substructure observable.

\begin{figure*}[t]
\centering
\includegraphics[scale=0.4]{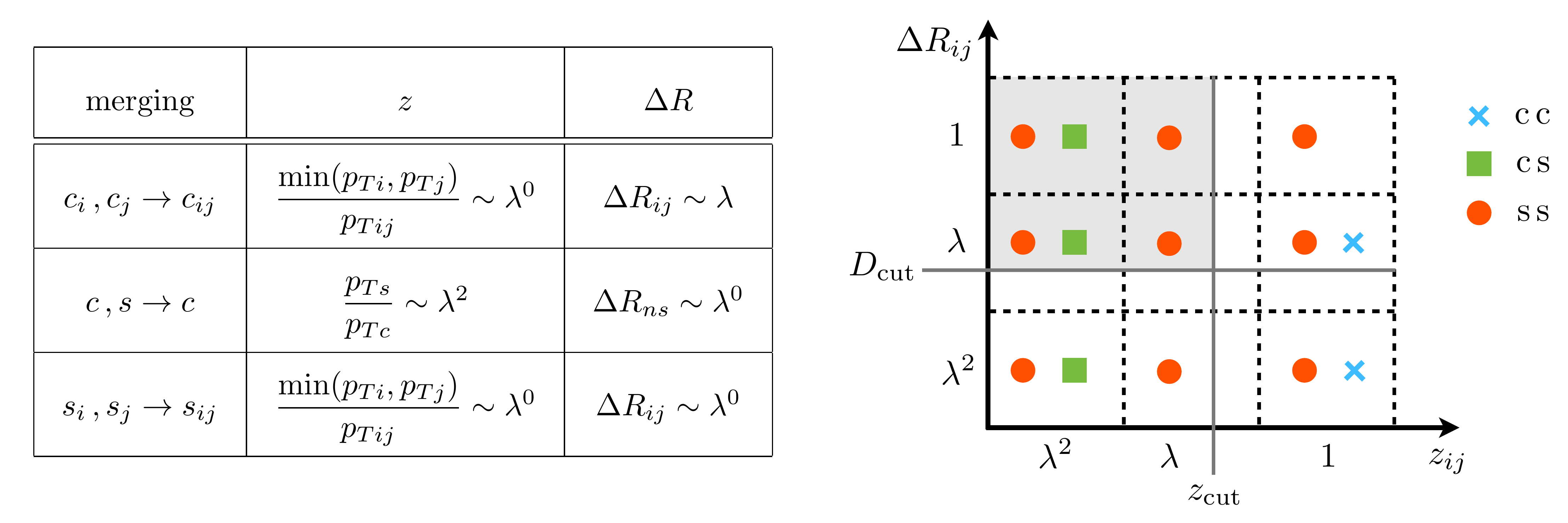}
\caption{The scaling of recombination variables $z$ and $\Delta R$ in the pruning algorithm.  Left: table of scaling for collinear-collinear, collinear-soft, and soft-soft recombinations.  Right: Plot of $\Delta R$ vs.~$z$, and the regions that each type of recombination can occupy and still give a leading power contribution to the cross section.  The shaded region is the region where pruning takes place, with the choice $z_{\rm cut}, D_{\rm cut} \sim \ord{\lambda}$.}
\label{fig:pruningrecombinations}
\end{figure*}

The requirement that collinear particles not be removed from the jet constrains the scaling of the parameters $z_{\rm cut}$ and $D_{\rm cut}$, since $c,c\to c$ recombinations cannot be pruned.  In \fig{pruningrecombinations}, we show the scaling of the kinematic variables $z$ and $\Delta R$ for the three kinds of recombinations. We include the contribution from all regions of phase space allowed in the effective theory that could contribute at leading power, including those with non-characteristic scaling (e.g., $\Delta R_{cc} \sim \ca{O}(\lambda^2)$) as discussed in \subsec{SCEToverview}. 

From the figure, we can see that no collinear subjets will be removed if $z_{\rm cut}$ is chosen to scale as
\be \label{eq:zcutscaling}
z_{\rm cut} \lesssim \ca{O}(\lambda) \,.
\ee
No scaling constraint needs to be made on $D_{\rm cut}$, since this is a cut on the relative angle between particles that depends entirely on either collinear (for $c,c\to c$ merging) or soft (for $c,s \to c$ or $s,s \to s$ merging) momenta.  With the scaling choice in~\eq{zcutscaling}, pruning satisfies the necessary condition for OPS factorization and could be factorized.  We note that the constraint on the scaling of $z_{\rm cut}$ is consistent with the intention of pruning to remove soft, wide angle radiation from the jet substructure.  The original choices for the parameters, $z_{\rm cut} = 0.10 - 0.15$ and $D_{\rm cut} = m_J/p_{TJ}$, are consistent with $z_{\rm cut}, D_{\rm cut} \sim \lambda$, which is depicted in the right side of \fig{pruningrecombinations}.

\subsubsection{Characteristic Behavior of Pruning}
\label{subsec:pruning}

Just as we did for jet algorithms, we can apply the ideas of power counting to study the behavior of pruning.  Using power counting, we can develop a picture of what remains in a jet after pruning for different reclustering algorithms (here we consider $\kt$, C/A, and anti-$\kt$).  Focusing on energetic jets with a single collinear sector, we find that this simple picture describes the jets remarkably well, as the qualitative picture agrees with the behavior of pruned jets in a Monte Carlo simulation.

We choose $z_{\rm cut}$ and $D_{\rm cut}$ to scale as $\lambda$, as in the standard pruning implementation.  We will make use of \fig{pruningrecombinations}, which plots the regions where each type of recombination can contribute to the cross section at leading power.  Two facts help us develop a picture of pruning that we can apply to different algorithms:
\begin{list}{\labelitemi}{\leftmargin=1.5em}
\item Factorization requires that no collinear subjet be pruned.
\item Every soft subjet will eventually be recombined with a collinear subjet, where it will be pruned unless $\Delta R_{ns} < D_{\rm cut}$.
\end{list}
The first item means that the full collinear sector will remain after pruning.  The second means that we can determine which soft particles remain after pruning by determining how unpruned soft-soft recombinations shape the soft phase space.  The relative ordering of soft-collinear and soft-soft recombinations in a reclustering algorithm will determine the soft phase space after pruning.  We refer to \fig{AlgorithmOrdering:metrics} and the discussion in \subsec{jetalgs} for the ordering of recombinations for the anti-$\kt$, C/A, and $\kt$ algorithms.

To determine the soft region that remains after pruning, we will consider the configuration shown in \fig{cssConfig} with the jet size $R$ replaced by $D_{\rm cut}$. The pair of soft particles, $s_1$ and $s_2$, have
\be
\Delta R_{ns_1} < D_{\rm cut} < \Delta R_{ns_2} \,.
\ee
If $s_1$ is merged with the jet before $s_1$ and $s_2$ are combined, then $s_2$ will be pruned.  Therefore we can determine when
\be \label{eq:metriccomparesoftPS}
\rho_{s_1\, s_2} < \rho_{n s_1} \,,
\ee
which will tell us the parametric size of the region of soft phase space that  remains after pruning.

The anti-$\kt$ algorithm characteristically merges soft-collinear pairs before soft-soft pairs.  The comparison in \eq{metriccomparesoftPS} for anti-$\kt$ is
\be
\Delta R_{s_1 \, s_2} < \frac{\min(p_{Ts_1}, p_{Ts_2})}{p_{Tc}} \Delta R_{ns_1} \ll D_{\rm cut} \,.
\ee
The ratio of soft and collinear $p_T$ requires $\Delta R_{s_1\, s_2} \ll \Delta R_{ns_1}$, and so the region of phase space where $s_2$ is not pruned is power suppressed.  This means that for anti-$\kt$, the soft phase space after pruning is simply a disk of radius $D_{\rm cut}$ centered on the jet axis up to power corrections.

\begin{figure*}[t]
\centering
\includegraphics[scale=0.6]{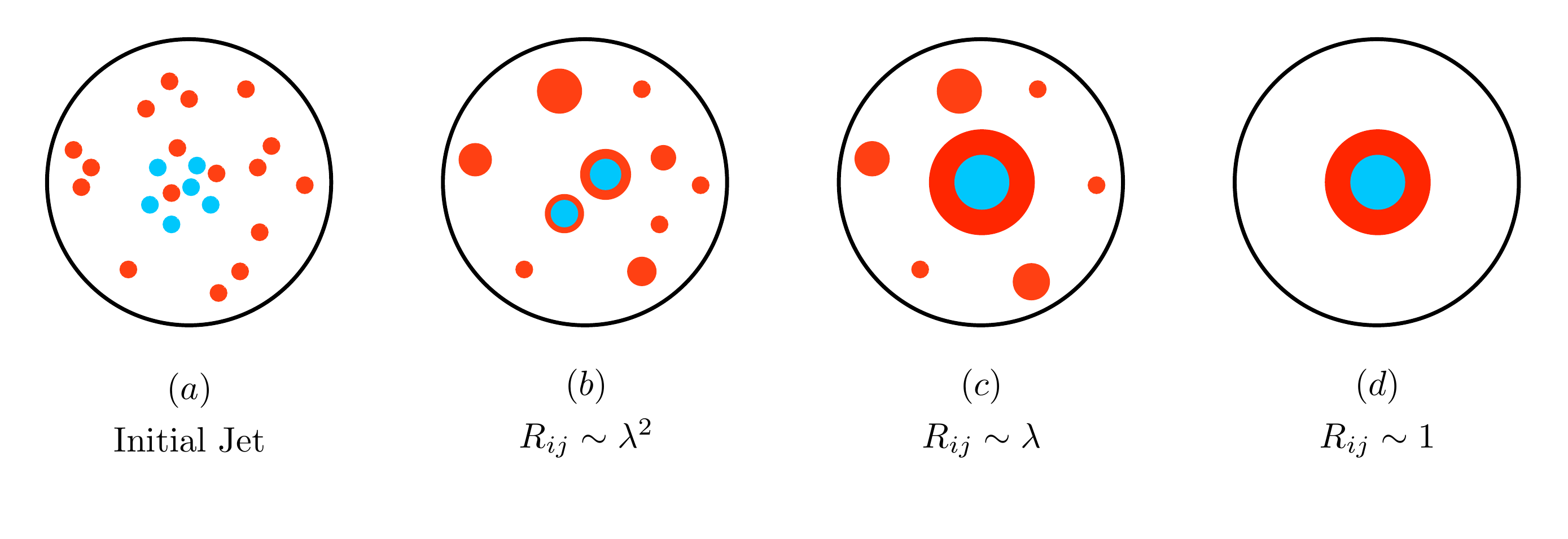}
\caption{The stages of pruning for C/A reclustering.  Groups of soft particles are red and groups of collinear particles are blue.  First, all pairs of particles at small angles are merged (a $\to$ b).  Then larger angle recombinations occur; collinear particles are merged together, and some pruning takes place (b $\to$ c).  Finally wide angle pairs of particles are merged, and any soft particles farther than $\approx 2D_{\rm cut}$ from the jet axis get pruned (c $\to$ d).}
\label{fig:pruningstepsCA}
\end{figure*}

The C/A algorithm characteristically merges soft-collinear and soft-soft pairs simultaneously.  The comparison in \eq{metriccomparesoftPS} for C/A is
\be
\Delta R_{s_1\, s_2} < \Delta R_{ns_1} < D_{\rm cut} \,.
\ee
This implies that soft particles within an angle $2D_{\rm cut}$ of the jet axis can remain after pruning.  Note that as the angle of $s_2$ to the jet axis grows, the recombined pair of soft particles are more likely to be farther than $D_{\rm cut}$ from the jet axis and be pruned.  Multiple soft-soft recombinations will mitigate this effect, and will tend to allow for wider angle soft particles to be merged into the jet.  However, the essential feature is that we expect the soft phase space to extend out to a radius $2D_{\rm cut}$, twice that for anti-$\kt$.  In \fig{pruningstepsCA}, we show the basic steps of pruning for C/A.

The $\kt$ algorithm characteristically merges soft-collinear and soft-soft pairs simultaneously, as in the C/A algorithm.  The comparison in \eq{metriccomparesoftPS} for $\kt$ is
\be
\Delta R_{s_1\, s_2} < \frac{p_{Ts_1}}{\min(p_{Ts_1},p_{Ts_2})} \Delta R_{ns_1} \,.
\ee
If $p_{Ts_2} > p_{Ts_1}$, then this constraint is the same as C/A.  But if $p_{Ts_2} < p_{Ts_1}$, then $s_2$ can be at wider angle to the jet axis and still be merged with $s_1$.  Additionally, since $s_2$ is softer in this case, it is more likely that the recombined pair of soft particles will be closer than $D_{\rm cut}$ from the jet axis.  This suggests that the soft phase space remaining after pruning will be a disk of radius $2D_{\rm cut}$ (as for C/A) plus a region that extends out to larger angles where a fraction of the soft particles are retained.

We can test this basic picture predicted by power counting by looking at the effect of pruning on jets from the Boost 2010 samples.  Using the same events as in \fig{algmerging}, we find jets with the anti-$\kt$ algorithm using $R = 1.0$.  Selecting jets with $m/p_T < 0.25$, we prune them using anti-$\kt$, C/A, and $\kt$ reclustering using the FastPrune v0.4.3 plugin for FastJet\footnote{Pruning and other substructure algorithms are natively implemented in FastJet 3.0.0, released as this work was being completed.}, choosing $D_{\rm cut} = 0.1$ and $z_{\rm cut} = 0.1$ \cite{FastPrune}.

\begin{figure*}[t]
\centering
\includegraphics[scale=0.31]{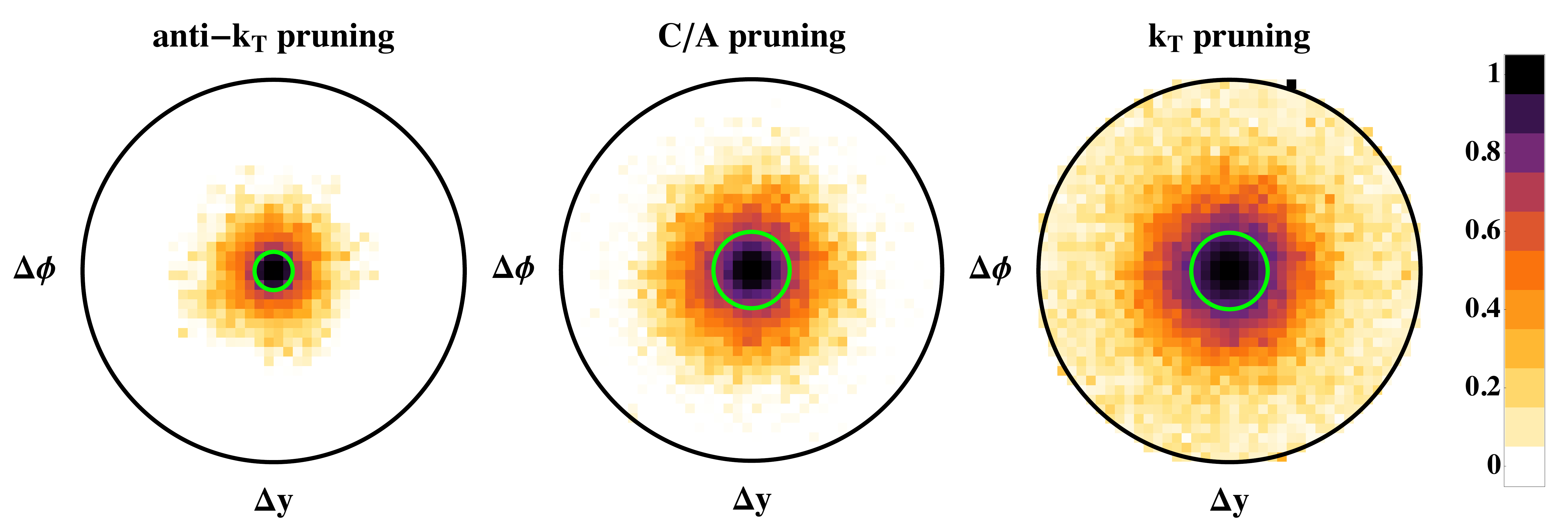}
\caption{The fraction of $p_T$ remaining after pruning as a function of the location in the jet, for an ensemble of jets from the Boost 2010 events described in the text.  The jets are found with anti-$\kt$ using $R = 1.0$ and have $p_T > 200$ GeV and $m/p_T < 0.25$.  Pruning was performed with anti-$\kt$, C/A, and $\kt$ reclustering, with $z_{\rm cut} = 0.10$ and $D_{\rm cut} = 0.1$.  In each bin of size 0.05-by-0.05 in $\Delta y$ and $\Delta \phi$, we plot the fraction of $p_T$ that remains after pruning.  The green circle shows the region predicted by power counting where most of the energy remains.  This circle has radius $D_{\rm cut}$ for anti-$\kt$ and $2D_{\rm cut}$ for C/A and $\kt$.}
\label{fig:pruningaveragealgs}
\end{figure*}

To quantify what remains after pruning, we bin all particles in their relative $\Delta y$ and $\Delta \phi$ to the jet axis.  For each bin, we sum over all jets and determine what fraction of $p_T$ remains after pruning.  This tells us how much $p_T$ remains after pruning as a function of the location in the jet.  For each algorithm, this is shown in \fig{pruningaveragealgs}.  We also show the region predicted by power counting where the soft phase space remains largely unpruned and most of the energy remains.  The Monte Carlo simulation gives excellent agreement with the basic predictions of power counting.

\subsection{Trimming}
\label{subsec:trimming}

The trimming algorithm is intended improve the resolution of jets coming from light partons by removing contamination from initial state radiation, the underlying event, and pileup \cite{Krohn:2009th}.  Trimming uses the fact that the energetic components of final state radiation in jets are well collimated and can be clustered into subjets by a jet algorithm with radius $R_{\rm sub} \ll R$.  Fat jets used to capture nearly all of the radiation from the evolution of an energetic particle can then be trimmed, and the mass resolution of reconstructed decays is subsequently improved.

Trimming defines subjets through a secondary jet algorithm.  Starting with jets found with large $R$, the trimming procedure is:
\begin{list}{\labelitemi}{\leftmargin=1.5em}
\item[1.] Recluster the jet with an algorithm with radius $R_{\rm sub} \ll R$.  The jets found by this algorithm are candidate subjets.
\item[2.] Remove any candidate subjets with $p_T < p_T^{\rm cut} \equiv f_{\rm cut} \Lambda_{\rm hard}$, where $f_{\rm cut}$ and $\Lambda_{\rm hard}$ are parameters.  The remaining subjets form the new, trimmed jet.
\end{list}
In the original trimming definition, $\Lambda_{\rm hard} \sim p_{TJ}$ or $\sqrt{\hat{s}}$, and $f_{\rm cut} \ll 1$ is a dimensionless parameter.

As discussed for pruning, factorization requires that a collinear subjet cannot be removed unless phase space constraints require that the subjet is well separated and can be treated as its own collinear sector.  This requires that $p_T^{\rm cut}$ must scale with $\lambda$ such that the cut does not remove collinear subjets:
\be
\frac{p_T^{\rm cut}}{p_{TJ}} \lesssim \lambda \,.
\ee
Any larger scaling would allow a region of phase space where collinear subjets are removed.  

No analogous constraint exists for the subjet radius $R_{\rm sub}$.  However, an important principle applies in choosing the value of $R_{\rm sub}$.  The restriction operator $\widehat{\ca{R}}$ implements the phase space cuts of the measurement.  These cuts determine the kinematics of the jets that count towards the cross section.  The cuts that substructure methods place tend to select for kinematics that resemble a heavy particle decay and look unlike QCD.  For instance, many substructure algorithms require well separated clusters of energy in the jet that look like subjets from a decay, strongly cutting on the QCD background.  The choice of $R_{\rm sub}$ will affect the resulting substructure, especially if additional constraints are placed on the collinear subjets.

When considering what types of observables have OPS factorization, we find a general constraint on jet substructure:
\begin{list}{\labelitemi}{\leftmargin=1.5em}
\item We cannot measure observables for individual subjets unless the subjets are well-separated in the jet.
\end{list}
Because soft particles do not resolve individual collinear particles, the soft constituents of an individual subjet are ambiguous.  This implies that we cannot measure certain properties of individual subjets and maintain factorization.  For example, the subjet mass depends on knowing the soft and collinear constituents of the subjet.  Although we know which soft and collinear modes are in the jet after trimming, we do not know the distribution of the soft modes inside the collinear subjets.  This implies that we can factorize jet shape observables which act on the whole (trimmed) jet, but not subjet observables that require knowing the soft constituents of particular subjets.  This constraint applies to other substructure algorithms and jet algorithms in general.

If we are interested in measuring more exclusive jet observables, we can impose additional constraints on the subjet clustering procedure.  If we require the collinear subjets to be well-separated, then we can treat each subjet as a distinct collinear sector.  This means separate collinear subjets would be equivalent to separate jets from the perspective of soft modes.  The soft function can know about the direction and $p_T$ of each subjet, and soft modes can be assigned to specific collinear subjets.  

It is fairly straightforward to require that each subjet be well-separated.  For instance, we can require that the collinear subjets be separated by an angle $R_{\rm sep} \gg R_{\rm sub}$.  While this will make trimming more complex, such constraints are necessary for factorization of jet observables which measure properties of individual subjets sensitive to soft momenta.
 
\subsection{The Mass-Drop Filter Method}
\label{subsec:filtering}

The mass-drop filter (MD-F) method is intended to tag boosted Higgs decays into a fat jet \cite{Butterworth:2008iy}.  The algorithm works by declustering a found jet, stepping backwards through the recombinations, until a declustering characteristic of a boosted decay is identified.  These objects are further declustered into subjets, and these subjets are filtered by removing the softer subjets.  Declustering and filtering remove contamination from underlying event and pileup, which has a characteristically lower energy scale than the hard radiation in the jet.  The initial declustering step significantly reduces the QCD background, while the filtering step improves the mass resolution of jets containing a boosted Higgs.

The MD-F method starts with jets defined by the C/A algorithm.  MD-F defines two subjets through a declustering procedure which steps back through the recombinations of the jet algorithm:
\begin{list}{\labelitemi}{\leftmargin=1.5em}
\item[1.] Label the jet $J$ and perform a single declustering step to obtain two candidate subjets $j_1$ and $j_2$ with $m_{j_1} > m_{j_2}$.
\item[2.] On the candidate subjets, test if
\begin{align}
&a \equiv \frac{m_{j_1}}{m_J} < \mu \quad \textrm{and} \nn \\
&y \equiv \frac{\min(p_{Tj_1}^2,p_{Tj_2}^2)}{m_J^2} \Delta R_{j_1,j_2}^2 > y_{\rm cut} \, .
\end{align}
\item[3.] If these cuts are passed, the splitting $J\to j_1,j_2$ is a hard declustering, and $j_1, j_2$ are the hard subjets; proceed to the next step.  If these cuts are not passed, then redefine the jet to be the heavier subjet $j_1$ and return to step 1.  Repeat until hard subjets are found.
\end{list}
The parameters $\mu$ and $y_{\rm cut}$ are inputs to the algorithm.  These subjets are then filtered:
\begin{list}{\labelitemi}{\leftmargin=1.5em}
\item[4.] Define $R_{\rm filt} = \min\left(0.3, {\sfrac12}R_{j_1,j_2}\right)$.  Decluster $j_1$ and $j_2$ down to $R_{\rm filt}$ and keep the 3 highest $p_T$ subjets from the declustering.  These 3 subjets define the new jet.
\end{list}
Filtering steps back through the recombinations of the C/A algorithm until the declustering angle is less than $R_{\rm filt}$, keeping only the highest $p_T$ subjets from the declustering.

The characteristic scaling of the variables $a$ and $y$ used in the declustering step are given in Table~\ref{table:MDFscaling}.  MD-F is applied to energetic jets and since declustering discards the lighter (softer) of the two subjets, the parent jet in declustering will always be collinear. 

The original choices for the parameters $\mu = 0.67$ and $y_{\rm cut} = 0.09$ are consistent with $\mu \sim \lambda^0$ and $y_{\rm cut} \sim \lambda$ \cite{Butterworth:2008iy}.
\renewcommand{\arraystretch}{2.5}
\begin{table*}[!ht]\normalsize
\caption{\normalsize{Scaling for the variables $a$ and $y$ in the MD-F algorithm.}}
\begin{center}
\begin{tabular}{|c|c|c|}
\hline
declustering & $a$ & $y$ \\
\hline
\hline
$c_J \to c_1\,, c_2$ & $\dfrac{m_{c_1}}{m_{c_J}} \sim \lambda^0 $ & $\dfrac{\min\left(p_{T c_1}^2, p_{T c_2}^2 \right)}{m_{c_J}^2} \Delta R_{c_1 , c_2}^2 \sim \lambda^0$ \\
[1ex]\hline
$c_J \to c_1\,, s_2$ & $\dfrac{m_{c_1}}{m_{c_J}} \sim \lambda^0$ & $\dfrac{p_{T s_2}^2}{m_{c_J}^2} \Delta R_{n , s_2}^2 \sim \lambda^2$ \\
[1ex]\hline
\end{tabular}
\end{center}
\label{table:MDFscaling}
\end{table*}%

The MD-F algorithm introduces many complications for factorization.  We first analyze the declustering procedure that determines the hard subjets, followed by the filtering procedure.

\subsubsection{Declustering}
\label{subsec:declustering}

The basic process of declustering requires knowledge of the ordering of recombinations in the jet.  However, as discussed in \subsec{general}, the relative ordering of recombinations in the soft and collinear sectors is not available in a factorizable cross section.  Therefore we cannot know which declustering step comes first: $c\to c,c$ or $c\to c,s$.  Unless additional kinematic constraints are placed on the substructure, MD-F and other algorithms that use declustering do not factorize.

Using insight from the previous examples, we discuss how a declustering method could be factorized.  If there are two well-separated collinear sectors in a single jet, then the first $c\to c,c$ declustering will resolve these collinear sectors.  In this case the two collinear subjets from the declustering can be resolved by the soft sector, as they are essentially separate jets.  This means that the kinematics of the $c\to c,c$ splitting can be known in the soft sector, and the constraints from merging order no longer apply.

The idea that the first $c\to c,c$ declustering define the hard subjets is physically sensible, since the kinematics resemble a hard splitting.  Furthermore, in the initial $c\to c,s$ splittings the soft particles will be at wide angles to the jet, which are the recombinations that most substructure methods aim to remove.  After discussing the filtering aspect of MD-F, we perform a power counting analysis on the declustering step of MD-F and suggest ways to require the jet to have multiple well-separated collinear sectors, using the kinematic cuts in MD-F.

\subsubsection{Filtering}
\label{subsec:filtering}

The filtering procedure also introduces complications for factorization.  The basic process of filtering is simple:
\begin{list}{\labelitemi}{\leftmargin=1.5em}
\item Decluster to a given scale and keep only the $N$ hardest subjets.
\end{list}
Without any additional kinematic constraints, this violates factorization.  The collection of $N$ subjets that is kept after filtering will have $N_{\rm coll}$ collinear subjets and $N - N_{\rm coll}$ soft subjets.  If $N_{\rm coll}^{\rm tot}$ is the total number of collinear subjets before the filtering step is applied, then
\be
N_{\rm coll} = \min(N, N_{\rm coll}^{\rm tot}) \,.
\ee
Unless $N_{\rm coll}$ is fixed by a kinematic constraint, the number of soft subjets that are removed by filtering depends on $N_{\rm coll}$, which is a phase space constraint coming from the collinear sector.  Since the soft function cannot know about this constraint, factorization is broken.

The similarity of trimming to the filtering step suggests a simple alternative.  Instead of a cut keeping the $N$ hardest subjets, if only subjets with a $p_T > p_T^{\rm cut}$ are kept, as in trimming, then if $p_T^{\rm cut} \sim \ca{O}(\lambda)$ only the collinear subjets will be kept and factorization can be preserved.  As with trimming, only jet shape observables that sum over all particles in the jet can be factorized. Observables such as subjet masses, where it is required to know which collinear subjet a soft particle is in, do not factorize without additional kinematic constraints on the subjets.

\subsubsection{Making MD-F Factorizable}
\label{subsec:MDFfactorizable}

A power counting analysis of the declustering step of MD-F can reveal simple ways to impose the kinematic constraint that the jet contain multiple well-separated collinear sectors.  The kinematic cuts of the MD-F algorithm make it natural for the first $c\to c,c$ declustering to define the hard subjets.  We only need to impose the constraint that the declustered collinear subjets are well-separated, i.e., that they are collimated compared to their separation, so that the collinear modes in each subjet do not overlap.  Furthermore, we must require that all $c\to c,s$ splittings fail the hard subjet cuts.

Consider the power counting of collinear-collinear and collinear-soft splittings in terms of $a$ and $y$, given in Table~\ref{table:MDFscaling}.  These are valid when there is one collinear sector in the jet.  If we choose
\be
y_{\rm cut} \sim \lambda \,,
\ee
then $c\to c,s$ splittings will always fail this cut, and $c\to c,c$ splittings will pass this cut.  Since $c\to c,c$  splittings have $a\sim \lambda^0$, unless $\mu \ge a_{\max}$ there is a region of phase space where $c\to c,c$ splittings will fail the cut on $a$.  However, we can use this cut to require that the jet have two well-separated collinear sectors.  If $\mu \sim \lambda$, then any splitting $J \to j_1, j_2$ with $a < \mu$ will have
\be
m_{j_2} < m_{j_1} \ll m_J \,.
\ee
This implies that the subjets are collimated compared to their separation.  A $c\to c,c$ declustering will only pass this cut if the two collinear subjets are well-separated, meaning we can treat them as separate collinear sectors.

Therefore, if we require
\be
\mu \sim \lambda \,, \quad y_{\rm cut} \sim \lambda \,,
\ee
then the declustering step of MD-F passes the power counting test for soft-collinear phase space factorization.  With this choice, a $c\to c,c$ declustering where the two collinear subjets are separate collinear sectors will pass the cuts.  Only jets with two (or more) collinear sectors will pass the cuts.  While $\mu \sim \lambda$ is more restrictive than $\mu\sim \lambda^0$, it is required for OPS factorization.  The hard subjets can subsequently be filtered using a factorizable procedure discussed above.

Reducing the value of $\mu$ places more restrictive cuts on the substructure, which will lower the Higgs tagging efficiency.  This is a general property of more exclusive factorization theorems -- we must be more exclusive in the jet substructure, to ensure that factorization constraints can be satisfied.  However, a lower signal efficiency will be accompanied by a lower background mistag rate, meaning the substructure method can still be effective.  For the MD-F tagger, we perform a simple Monte Carlo study to determine the impact of choosing a smaller $\mu$.  We generate $Zj$ and $ZH$ events with $m_H = 115$ GeV using Pythia v8.145 \cite{Sjostrand:2007gs}, and select jets with $p_T$ between 200 and 300 GeV, decaying the $Z$ leptonically.  We carry out the MD-F algorithm with the default value $y_{\text{cut}} = 0.09$ and two $\mu$ values, 0.67 (the default value) and 0.25 (the small value).  

\begin{figure*}[th!]
\includegraphics[scale=0.48]{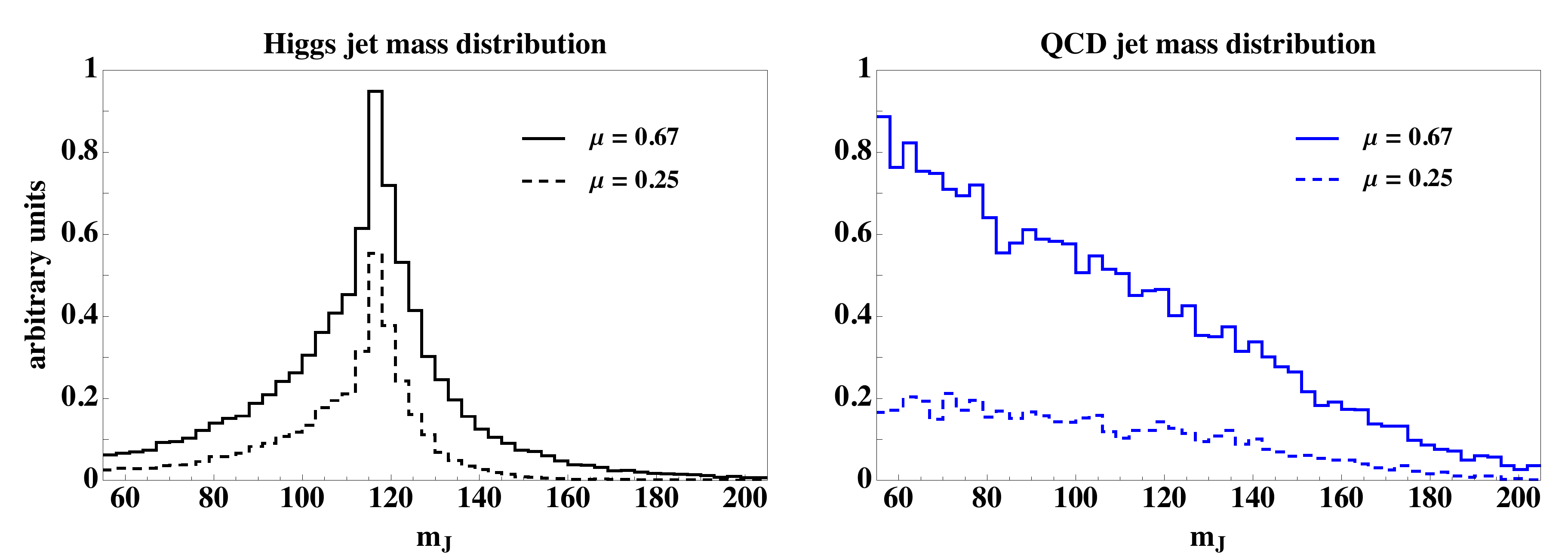}
\caption{The Higgs mass distribution (left) and background jet mass distribution (right) for the MD-F algorithm with the default $\mu$ (0.67) and small $\mu$ (0.25).  The small $\mu$ allows for factorization.  With small $\mu$ the signal efficiency is lower, but there is a compensating drop in the mistag rate.} \label{fig:MDF}
\end{figure*}

In \fig{MDF}, we plot the mass distributions for the Higgs signal and the background, for both values of $\mu$.  We can see that while the overall tagging efficiency is reduced, the QCD background significantly decreases.  Choosing a mass range for the Higgs candidate of $[100,130]$ GeV, we find that with a small $\mu$ the tagging efficiency is reduced to 48\% of the efficiency with the default parameters, with the signal purity (signal over background, $S/B$) increases by a factor of 1.8 and the significance (signal over noise, $S/\sqrt{B}$) decreases by a factor of 0.94.  We note that choosing a more restrictive value of $\mu$ will further reduce the tagging efficiency.  The significance is nearly unchanged while the lower efficiency is accompanied by a higher signal purity.

\subsection{N-subjettiness}
\label{subsec:Nsubjettiness}

$N$-subjettiness is a jet shape constructed analogously to $N$-jettiness, an event shape \cite{Stewart:2010tn}.  $N$-jettiness provides a veto on additional jets, effectively characterizing how $\le \! N$-jet-like an event is ($N$ jettiness does not veto against fewer jets -- if $N$-jettiness is small, so is $M$-jettiness for $M < N$).  $N$-subjettiness is a natural generalization of this.  It provides a veto against additional subjets through a measure $\tau_N^{(\beta)}$ \cite{Thaler:2010tr,Thaler:2011gf}, defined as
\be \label{eq:NsubDef}
\tau_N^{(\beta)} \equiv \min_{\{n_J\}} \frac{1}{d_0} \sum_i p_{Ti} \min \left\{ \left(\Delta R_{1,i}\right)^{\beta}, \ldots, \left(\Delta R_{N,i}\right)^{\beta} \right\} \,,
\ee
where the measure $\Delta R_{J,i}$ is the separation between a particle $i$ and subjet $J$.  The inner minimization picks the smallest $\Delta R_{J,i}$ over all $N$ subjets $J$, while the outer minimization chooses the subjet axes $n_J$ that minimize the sum.  The normalization factor $d_0$ is
\be
d_0 \equiv \sum_i p_{Ti} R^{\beta} \,,
\ee
with $R$ the jet radius.  The power $\beta$ is required to be non-negative by collinear safety.  

When $N$-subjettiness is small, the jet is $\le \! N$-subjet-like.  A large value of $\tau_N$ means there are more subjets in the jet, which provides the veto against additional subjets.  Because of this, the ratio $\tau_{N}/\tau_{N-1}$ is a useful discriminant to identify boosted $N$-body decays.  This ratio will be small for the $N$-subjet-like boosted decay, since $\tau_N$ will be small and $\tau_{N-1}$ will be large.  For QCD, however, we expect this ratio to be $\ca{O}(1)$ except for 3 (or more) subjet-like jets, which are rarer in QCD.  When compared to other techniques using the Boost 2010 Monte Carlo samples, this single variable provides a more effective top tagger over nearly the entire range of top tagging efficiency \cite{Thaler:2011gf}.  We now discuss factorization considerations for jet shapes in general, and some of the subtleties that arise with $N$-subjettiness.

Most shapes (like jet mass or $N$-subjettiness) are linear in the final state momenta, taking the form
\be
\tau = \sum_{i \in J} f_{\tau} (k_i) \,,
\ee
where collinear safety requires $f_{\tau} (k) \propto (E_k)^1$.  For these linear jet shapes, the observable splits into separate sums of collinear and soft contributions,
\be
\tau = \tau_c + \tau_s = \sum_{c_i \in J} f_{\tau} (k^c_i) + \sum_{s_i \in J} f_{\tau} (k^s_i) \,.
\ee
For simple functions $f_{\tau}$, this makes OPS factorization straightforward.  There are separate contributions for soft and collinear modes, and the only phase space constraints that can spoil factorization are in the algorithm used to find the jets.  

For non-linear jet shapes, such as jet mass, one must ensure that the soft and collinear contributions separate.  Typically this involves showing that the soft contributions depend only on the total jet momentum instead of the momentum of individual collinear particles.  For instance, for collinear-soft pairs their contribution to the jet mass is
\be
\sum_{c,s \in J} E_c \, E_s \, \Delta R_{ns}^2 = E_J \sum_{s\in J} E_s \, \Delta R_{ns}^2 \,.
\ee
This gives a contribution like $m_s^2$ in \eq{Mjet} in the case of the hemisphere jet mass.

We note that factorization for jet shapes is generally limited to some regime of the observable where soft and collinear modes dominate the contribution to the observable.  For example, soft and collinear dynamics describe the regime of small jet mass, $m_J \ll E_J$.  In the large jet mass regime, $m_J \sim E_J$, hard modes contribute and perturbation theory is sufficient to accurately calculate the mass distribution.  For $N$-jettiness, the region of small $\tau_N$ contains large logarithms of $\tau_N$ in the perturbative series, and factorization is needed to resum these logs.  In the region of large $\tau_N$, factorization is not needed and perturbation theory can be used.

A power counting analysis of the jet shape observable can be useful in determining if the soft and collinear modes of \SCETI provide the relevant description of the final state. This addresses the first key step in the proof of factorization outlined in \subsec{factoverview}\,. We carry out this analysis for $N$-subjettiness here, which follows directly from the study of angularities \cite{Berger:2003iw,Berger:2003pk} in the effective theory \cite{Bauer:2008dt,Hornig:2009vb,Ellis:2009wj,Ellis:2010rwa}.  For $\beta = (2-a)$ $N$-subjettiness has similar kinematic dependence to the angularity jet shape $\tau_a$. We can look at the contributions of a soft and collinear particle to $N$-subjettiness:
\begin{align}
\text{collinear: }& \tau_N^{(\beta)\,c} = \frac{1}{d_0} \, p_{T}^c \, (\Delta R_{J,c})^\beta \sim \lambda^{\beta}, \nn \\
\text{soft: }& \tau_N^{(\beta)\,s} = \frac{1}{d_0} \, p_{T}^s \, (\Delta R_{J,s})^\beta\sim \lambda^2 \,.
\end{align}
The relationships between the power counting parameter and the jet and soft scales determines $\mu_{J,S}$ in terms of the observable, $\tau_N^{(\beta)}$:
\begin{align}
\mu_J &\sim E_{ J} \lambda \quad \Rightarrow \quad \mu_J \sim E_J \left(\tau_N^{(\beta)} \right)^{1/\beta} \,, \nn \\
\mu_S &\sim E_{J}\lambda^2 \quad \!\!\! \Rightarrow \quad \mu_S \sim E_J \, \tau_N^{(\beta)} \,.
\end{align}
For $\beta>1$ the hard, jet, and soft scales are well separated,
\be
E_J \, \tau_N^{(\beta)} \ll E_J \left(\tau_N^{(\beta)} \right)^{1/\beta} \ll E_J \,,
\ee
and \SCETI is a useful effective theory description of the small $N$-subjettiness region, $\tau_N^{(\beta)} \ll 1$. When $\beta=1$, $\mu_J = \mu_S$ and therefore the soft and collinear modes of \SCETI do not provide the correct description and a different EFT is required. In this case \SCETII, with soft modes which scale as $E_J (\lambda, \lambda, \lambda)$, provides the appropriate description of $N$-subjettiness. This corresponds to the angularity $a=1$, known as jet broadening \cite{Catani:1992jc, Dokshitzer:1998kz, Chiu:2011qc, Becher:2011pf}.

$N$-subjettiness is defined with respect to a projection of soft and collinear momenta along subjet axes. OPS factorization for $N$-subjettiness in \SCETI requires that 
\begin{list}{\labelitemi}{\leftmargin=1.5em}
\item Subjet axes obtained by minimizing $N$-subjettiness must be independent of the soft momenta in the jet, up to power corrections.
\end{list}
This says that if $N$-subjettiness OPS factorizes in \SCETI  (the appropriate theory for $\beta>1$) the minimization over $n_J$ in \eq{NsubDef} is determined by the sum over the collinear particles in each subjet. The soft contribution only changes the jet axis by a small angle, which gives a power suppressed contribution to $\tau_N^{(\beta)}$. 
For $\beta=1$ the subjet axes are different from the sum of collinear particles in the subjet by an $\ca{O}(1)$ amount, \cite{Thaler:2011gf}. The contribution of the soft momenta of \SCETII must be considered and in the case of jet broadening this is taken in to account by a recoil term \cite{Dokshitzer:1998kz,Becher:2011pf}.  In \cite{Thaler:2011gf}, the authors discuss several ways to determine the subjet axes, and it would be interesting to determine whether the various choices allow for factorization.

In $N$-subjettiness, each subjet axis is associated with a subjet, made up of those particles whose contribution to $\tau_N$ comes from that subjet.  In a regime where $\tau_N$ is small, each subjet's contribution to $\tau_N$ is also small.  This implies that the clusters of collinear particles in each subjet are well separated, and we can treat them as distinct collinear sectors.  Therefore the soft particles can resolve the different subjets, and we can completely determine the constituents of each subjet.

\section{Beyond Soft-Collinear Factorization}
\label{sec:caveats}

The analysis we have presented applies to a wide swath of jet physics applications where soft and collinear modes dominate observables of interest and factorization of jet substructure algorithms is desired.  In this work we have discussed factorization constraints in the context of SCET, which provides a rigorous and powerful framework for factorization and resummation.  We stress that the underlying principles of SCET are no different than QCD, and factorization in SCET simply frames factorization in QCD in an effective theory picture.  This effective theory picture changes depending on what kind of observables we are studying, and so we discuss cases where the OPS factorization analysis must be considered more carefully.

First, we note that for jet physics applications with no large logarithms in the perturbative series, resummation may not be necessary to obtain accurate predictions\footnote{Even when logarithms are not very large, resummation can be helpful in reducing theoretical uncertainties and improving convergence.  This is often the case when dealing with logarithms of scale ratios in a regime where the scales are transitioning from being disparate to being of the same order.}.  In this case a fixed order calculation would be sufficient and SCET would not apply.  As we have seen, this is not the case for most jet substructure methods.  The presence of small parameters often guarantees large logarithms in the perturbative series, requiring resummation.

When resummation is required, considerations of the accuracy needed in the resummed prediction are important.  Often, a leading log (LL) or next-to-leading log (NLL) calculation has sufficient accuracy to compare to data, and other parts of the cross section (such as non-perturbative corrections) can have a larger error than the perturbative uncertainty.  In these cases an all-orders, all-logs factorization is more powerful than is needed, and the constraints of OPS factorization may be relaxed.  For example, the violations to factorization from the declustering and filtering steps may only occur at the next-to-next-to-leading log (NNLL) level.  If NLL resummation is all that is desired, then NNLL violations to factorization are not relevant to the calculability of the substructure algorithm.  In general, determining the order of violations to factorization is challenging, unlike the OPS factorization which is straightforward, but it would be interesting to determine this order for declustering and filtering.  We note that a study on the non-global logarithms in the MD-F algorithm suggests that violations to factorization do not appear at NLL \cite{Rubin:2010fc}.

There are many experimentally interesting jet observables that introduce additional scales beyond the simple picture in \fig{Scales}(b).  For example, nearly all jet substructure methods introduce small parameters that will generate additional large logarithms in a perturbative series. This complicates both the proof and the structure of the factorization theorem, and can reduce accuracy of predictions.  It requires either an extension of \SCETI with new modes that arise due to the presence of additional scales or an entirely different EFT.  For example, $N$-subjettiness with $\beta = 1$ is a case where \SCETII is the appropriate effective theory instead of \SCETI.  As we have seen in \subsec{Nsubjettiness}, a power counting analysis of the scale dependence of the soft and collinear modes can determine which effective theory is required.  In cases where an effective theory different from \SCETI is used, the first of the four key steps in the proof of factorization described in \subsec{factoverview} must be carefully considered, namely whether the soft and collinear modes of \SCETI contribute to the jet observable.

Certain jet physics applications may require a new effective theory description, one containing additional modes.  These go beyond the soft and collinear modes in \eq{scScales}.  There is active work on jet physics problems of this type \cite{Ellis:2009wj,Ellis:2010rwa,Kelley:2011ng,Hornig:2011iu, Bauer:2011uc,Kelley:2011tj,Hornig:2011tg}.  While the factorization theorems for these applications are undoubtedly more complicated, soft-collinear factorization is still at the heart of these problems.  The analysis we presented in this paper is still relevant and in fact the presence of additional modes makes the constraints on soft-collinear phase space factorization more restrictive.

\begin{figure}[t]
\centerline{ \scalebox{0.3}{\includegraphics{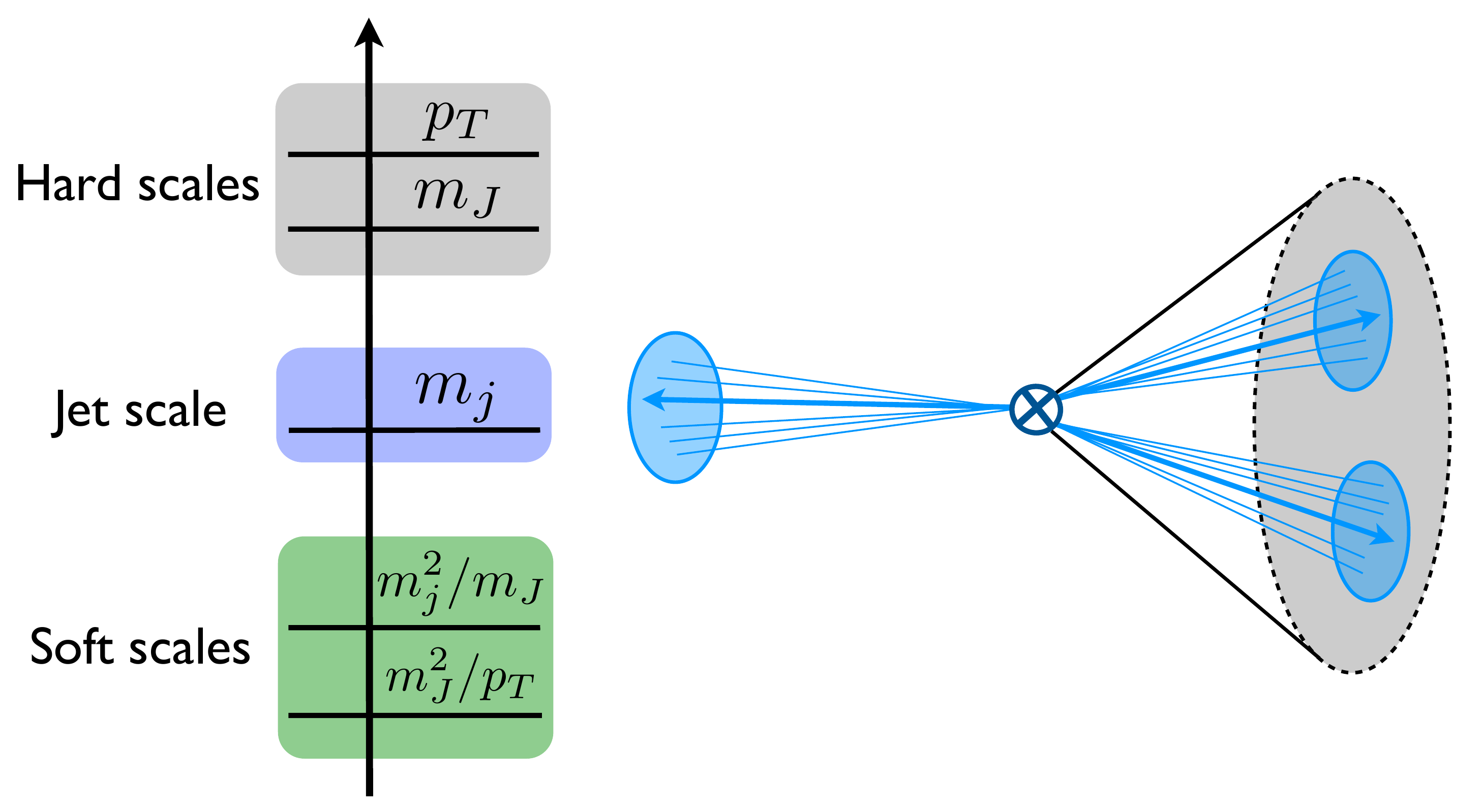}} }
\caption{Hierarchy of scales for two near-by jets relevant for jet substructure applications.}
\label{fig:Scales2}
\end{figure}

We consider an example, relevant for jet substructure, where an extension of \SCETI is needed.  Additional scales are introduced by the observable, and to deal with them we use \SCETp \cite{Bauer:2011uc}.  This is an EFT constructed to describe the configurations of the type depicted in \fig{Scales}(b) for $m_j \ll m_{J} \ll p_T$, where $m_j$ is the subjet mass and $m_J$ is the invariant mass of the two nearby subjets, which is of order of the mass of the decaying boosted object.  \SCETp is an extension of \SCETI with a csoft (collinear-soft) mode.  The modes in this theory have scaling 
\begin{align} \label{eq:scScales2}
\textrm{collinear: } & p_c  \sim p_T \left(1, \frac{m_j^2}{p_T^2}, \frac{m_j}{p_T} \right) \,, \nn \\
\textrm{soft: } & p_s \sim p_T \left(\frac{m_j^2}{p_T^2}, \frac{m_j^2}{p_T^2}, \frac{m_j^2}{p_T^2} \right)  \,, \nn \\
\textrm{csoft: } & p_{cs} \sim p_T \frac{m_j^2}{m_{J}^2} \left(1, \frac{m_{J}^2}{p_T^2}, \frac{m_{J}}{p_T} \right)  \,.
\end{align}
The nearby subjets lead to a large logarithm of the pairwise invariant mass, $\ln m_J / p_T$, in the perturbative series.  Without the csoft mode, this log cannot be resummed.  If we try to describe the two subjets with one collinear sector then we will not be able to simultaneously sum the large logs of the jet mass, $m_J$, and subjet mass, $m_j$.  The cross section for $pp\to N$ jets takes the schematic form
\begin{align}
\sigma_{N}
&= H_+(m_J)\, H_{N-1}(p_T) \, \Bigl[ B_a(m_j) B_b(m_j) \prod_{i = 1}^N J_i(m_j) \Bigr]
\nn \\
&\quad \otimes S_+(m_j^2/m_J) \otimes S_{N-1}(m_j^2/p_T)
\,,\end{align}
where the hard function $H_N$ and soft function $S_N$ of \eq{SCETfact} have been further factorized in to two separate functions that each depend on a single scale.  Despite the additional modes, the separation of soft and collinear phase space constraints is still a necessary condition for factorization of jet and soft functions for observables in \SCETp. 

It is important to note that this factorization requires the collinear subjets to be well-separated, e.g., $m_j \ll m_J$, as shown in \sec{jetsub}.  The utility of most substructure methods also relies on this separation, since if the boost of the heavy particle is too large, the subjets begin to overlap and can no longer be separately reconstructed. In the EFT this is reflected by the fact that they are no longer described by two separate jet functions, leading to an important factorization constraint for jet substructure methods.

Aside from cases where a new effective theory is required, in many jet physics applications techniques do not exist to resum some of the large logs in the perturbative series.  This is an ongoing area of study, and \SCETp is an example where a new technique to resum classes of logarithms (in this case when two collinear sectors become close) was developed.  One of the main challenges in this field is non-global logarithms \cite{Dasgupta:2001sh}.  Non-global logs arise from measurements over a restricted region of phase space, and are relevant for nearly every jet observable at the LHC.  These logs start at $\alpha^2 \ln^2$, and spoil even NLL resummation (in the exponent).  Resummation of the leading non-global logs exists only in the large $N_c$ limit, which is analytically quite challenging \cite{Banfi:2002hw}.  Interest in understanding these logs from the effective theory perspective is ongoing \cite{Kelley:2011ng,Hornig:2011iu,Hornig:2011tg}.  Beyond non-global logs, it is currently not known how to sum logarithms of parameters in the jet algorithm such as the jet or subjet radius (which can be small for substructure algorithms) or the $p_T$ cut used to veto additional jets.  Similarly, small parameters in jet substructure methods may need additional resummation.  As jet substructure matures, improvements in these areas seems likely.

\section{Conclusions}
\label{sec:conclusions}

We have explored factorization constraints for jet substructure algorithms.  Factorization plays a key role in being able to calculate and predict jet observables, and the constraints from factorization provide an important theoretical input for substructure.  SCET gives a framework to explore factorization, and OPS factorization is a key step.  OPS factorization is the physical statement that the phase space constraints on soft and collinear particles are independent at all orders in perturbation theory.  It is a requirement of observables that factorize in SCET, and provides strong constraints on jet observables, especially more exclusive ones like jet substructure.  Additionally, the power counting analysis that SCET provides is a useful tool to characterize the behavior of jet algorithms and substructure.

After discussing the factorization of well known jet algorithms and using power counting to study their behavior, we examined four substructure algorithms in \sec{jetsub}: the mass-drop filter (MD-F) algorithm, pruning, trimming, and $N$-subjettiness.  MD-F uses to basic steps common to nearly all substructure algorithms: declustering and filtering.  Declustering steps back through recombinations, while filtering removes soft subjets.  Both of these methods violate factorization, but a simple modification of the phase space cuts can specialize to a kinematic regime where factorization is allowed.  We studied the numerical impact of these modified cuts on Higgs tagging using a Monte Carlo study.  Pruning and trimming recluster the jet and make kinematic cuts to remove wide angle, soft radiation.  Each works uniquely and a simple power counting analysis reveals that factorization requires a certain scaling for parameters in each algorithm; this scaling is consistent with the original implementations.  We were able to use power counting to develop a simple picture of the remaining phase space of a pruned jet, and we found good agreement with this picture in a Monte Carlo study.  The factorization considerations for $N$-subjettiness, and jet shapes in general, are simpler.  Jet shapes measure a specific observable useful for substructure on all of the constituents of the jet, instead of modifying the substructure.  We discussed some of the subtleties of OPS factorization for more complex shapes like $N$-subjettiness.

As jet substructure methods are further explored both theoretically and experimentally, constraints from factorization and the power counting analysis that SCET provides are useful analysis tools.  Theoretical calculations of jet substructure will play an important role in understanding the data and the modeling of the Monte Carlo, and theoretically calculable substructure methods are needed.  The future of jet substructure promises to give many exciting results, and a deeper understanding can help probe the nature of new physics at the LHC.

\begin{acknowledgments}
We would like to thank Frank Tackmann and Jesse Thaler for helpful comments on a draft of this paper. We would also like to thank Christian Bauer, Nicholas Dunn, Andrew Hornig, Gavin Salam, and Iain Stewart for useful conversations.  We thank the Department of Energy's Institute for Nuclear Theory at the University of Washington for its hospitality and partial support during the completion of this work.
This work was supported in part by the Office of High Energy Physics of the U.S.\ Department of Energy under the Contract DE- AC02-05CH11231.  JW was supported in part by a LHC Theory Initiative Postdoctoral Fellowship, under the National Science Foundation grant PHY-0705682.
\end{acknowledgments}

\bibliographystyle{physrev4}
\bibliography{jets}

\end{document}